\documentclass[aps,pra,twocolumn,superscriptaddress,nofootinbib,amsmath,amssymb,floats,floatfix,english,showpacs,10pt]{revtex4-2}
\usepackage{amsmath,amssymb,graphicx}
\usepackage{hyperref}
\usepackage{dsfont}
\usepackage{dcolumn}
\usepackage{bm}
\usepackage{latexsym,epsfig}
\usepackage{verbatim}
\usepackage{comment}
\usepackage{stmaryrd}
\usepackage{color}
\usepackage{caption}
\usepackage{subcaption}
\usepackage{tikz}
\usepackage{siunitx}
\usetikzlibrary{shapes.geometric,positioning,fit,backgrounds}
\usepackage[normalem]{ulem}
\usepackage{todonotes}

\newcommand{\ignore}[1]{}

\begin{document}
\title{Applications of the Quantum Phase Difference Estimation Algorithm to the Excitation Energies in Spin Systems on a NISQ Device}

\author{Boni Paul}
\affiliation{Centre for Quantum Engineering Research and Education,
TCG Centres for Research and Education in Science and Technology,
Sector V, Salt Lake, Kolkata 700091, India}
\affiliation{Department of Physics, Indian Institute of Technology Tirupati, Yerpedu, Andhra Pradesh 517619, India. 
}

\author{Sudhindu Bikash Mandal}
\affiliation{Centre for Quantum Engineering Research and Education,
TCG Centres for Research and Education in Science and Technology,
Sector V, Salt Lake, Kolkata 700091, India}

\author{Kenji Sugisaki}
\affiliation{Centre for Quantum Engineering Research and Education,
TCG Centres for Research and Education in Science and Technology,
Sector V, Salt Lake, Kolkata 700091, India}
\affiliation{ Graduate School of Science and Technology, Keio University, 7-1 Shinkawasaki, Saiwai-ku, Kawasaki, Kanagawa 212-0032, Japan}
\affiliation{Quantum Computing Center, Keio University, 3-14-1 Hiyoshi, Kohoku-ku, Yokohama, Kanagawa 223-8522, Japan}
\affiliation{Keio University Sustainable Quantum Artificial Intelligence Center (KSQAIC), Keio University, 2-15-45 Mita, Minato-ku, Tokyo 108-8345, Japan}

\author{B. P. Das}
\affiliation{Centre for Quantum Engineering Research and Education,
TCG Centres for Research and Education in Science and Technology,
Sector V, Salt Lake, Kolkata 700091, India}
\affiliation{Department of Physics, School of Science, Tokyo Institute of Technology, Ookayama, Meguro-ku, Tokyo 152-8550,
Japan}

\begin{abstract}
The Quantum Phase Difference Estimation (QPDE) algorithm, as an extension of the Quantum Phase Estimation (QPE), is a quantum algorithm designed to compute the differences of two eigenvalues of a unitary operator by exploiting the quantum superposition of two eigenstates. Unlike QPE, QPDE is free of controlled-unitary operations, and is suitable for calculations on noisy intermediate-scale quantum (NISQ) devices. We present the implementation and verification of a novel early fault-tolerant QPDE algorithm for determining energy gaps across diverse spin system configurations using NISQ devices. The algorithm is applied to the systems described by two and three-spin Heisenberg Hamiltonians with different geometric arrangements and coupling strengths, including symmetric, asymmetric, spin-frustrated, and non-frustrated configurations. By leveraging the match gate-like structure of the time evolution operator of Heisenberg Hamiltonian, we achieve constant-depth quantum circuits suitable for NISQ hardware implementation. Our results on IBM quantum processors show remarkable accuracy ranging from 85\%  to 93\%, demonstrating excellent agreement with classical calculations even in the presence of hardware noise. The methodology incorporates sophisticated quantum noise suppression techniques, including Pauli Twirling and Dynamical Decoupling, and employs an adaptive framework. Our findings demonstrate the practical viability of the QPDE algorithm for quantum many-body simulations on current NISQ hardware, establishing a robust framework for future applications.
\end{abstract}
\maketitle
\section{Introduction}

Quantum computing has emerged as a transformative field with the potential to revolutionize approaches across various domains, including secure communications \cite{Grasselli2021}, financial modeling \cite{Orus2019}, machine learning \cite{Schuld2018}, combinatorial optimization \cite{Sanders2020}, and materials and drug discovery \cite{Cao2018}. One of its key strengths lies in its ability to address problems that are intractable for classical computers, especially those involving quantum many-body systems.
Classical computational techniques often face significant challenges due to the exponential growth in complexity, especially when trying to calculate larger and strongly correlated systems, while quantum computers are expected to handle them efficiently \cite{Feynman1982}.

Researchers have developed various quantum algorithms to effectively determine energy eigenvalues within quantum systems. Two notable techniques are the Variational Quantum Eigensolver (VQE) \cite{Peruzzo2014,Yung2014} and Quantum Phase Estimation (QPE) \cite{Abrams1999,Kitaev1995,NielsenChuang2010}. VQE employs a combination of quantum and classical computing principles, optimizing variational parameters in a parameterised quantum circuit to minimize the energy expectation value to find the ground state, which is well-suited for today's available noisy intermediate-scale quantum (NISQ) \cite{Preskill2018} devices.
However, VQE also encounters challenges like complex optimization with barren plateaus, rapid increase in shot count to reduce statistical error, reliance on classical optimizers, increased error rates with deeper circuits, difficulty in designing an efficient ansatz, and so on \cite{Tilly2022}. Conversely, fault-tolerant quantum computers (FTQC) are often seen as a distant possibility. QPE as an FTQC algorithm is expected to provide a more accurate eigenvalue estimate but it requires deep and complex quantum circuits. This complexity renders QPE demanding in terms of resources and prone to errors with the quantum devices currently available. Additionally, the process of calculating energy differences between quantum states typically requires multiple executions for each state in QPE or VQE, amplifying the computational efforts and highlighting the necessity for quantum algorithms that can more efficiently compute energy gaps with reduced resource usage. We emphasize that the estimation of energy differences between two eigenstates is an important task, especially in chemistry, because the information we can extract from experiments (e.g., spectroscopic measurements) is usually the energy differences, such as excitation energies, ionization energies, and reaction energy barriers, rather than the total energy itself.

 A Quantum Phase Difference Estimation (QPDE) \cite{Sugisaki2021} algorithm proposed by one of the authors of this paper addresses these challenges by offering a more efficient alternative for the energy gap estimation. It is designed to compute the difference between two eigenvalues of a unitary operator directly, using the quantum superposition of two eigenstates, and it is free of controlled-unitary operations. The latter feature is important especially when executing the algorithm on hardware with limited connectivity, while crosstalk errors further exacerbate implementation challenges. It should be noted that the control-unitary free implementations of the QPE algorithms have also been discussed in recent papers. \cite{Lin2022, Clinton2024}
 As quantum hardware progresses, particularly with developments in FTQC, algorithms like QPDE will become increasingly important. QPDE offers a scalable and practical solution for quantum simulations on both immediate experimental applications and long-term use in large-scale quantum simulations.
 
 While much of the existing research has focused on using this QPDE algorithm for challenges in quantum chemistry, such as vertical ionization energies, singlet--triplet energy gaps, and vertical excitation energies \cite{Sugisaki2021}, total energies \cite{Sugisaki2021b}, and numerical energy gradients \cite{Sugisaki2022}. It also shows promise in atomic and molecular physics for fine structure splitting calculations \cite{Sugisaki2023}. Recently, QPDE demonstration using superconducting quantum processor has also been reported \cite{Kanno2024}, with the aid of MPO-based classical compression of the quantum circuit and error suppression module provided by Q-CTRL \cite{Q-CTRL}. Moreover, spin models, while relatively simpler, serve as an excellent testing ground for this quantum algorithm on real quantum devices, offering important benchmarks for future studies involving more complex quantum many-body systems. 
Spin systems represent a foundational class of quantum many-body models that play a pivotal role in understanding a wide array of physical phenomena. These systems exhibit a diverse range of behaviours, including quantum coherence, magnetic ordering, phase transitions, and frustration. By applying the QPDE algorithm to spin systems, it becomes possible to perform highly accurate energy gap calculations, yielding deeper insights into their quantum characteristics. 

\begin{figure*}[t]
    \centering
    \begin{tikzpicture}[
        scale=0.75,
        every node/.style={transform shape},
        spin/.style={circle, draw, thick, minimum size=1.0cm, font=\large},
        coupling/.style={thick},
        label/.style={font=\large},
        box/.style={draw=black!30, rounded corners, inner sep=15pt},
        systemlabel/.style={font=\large\bfseries, above=0.5cm}
    ]
        \begin{scope}[xshift=1.0cm,local bounding box=two_spin]
            \node[spin] (1a) at (0,0) {1};
            \node[spin] (2a) at (3,0) {2};
            \draw[coupling] (1a) -- node[above, label] {$J_{12}=1$} (2a);
            
            \node[label, align=center] at (1.5,-1.5) {
                $E_T = -\frac{1}{2}$, \quad $E_S = \frac{3}{2}$ \\
                $E_S - E_T = 2.0$
            };
            
            \node[systemlabel] at (1.5,1) {Two-Spin System};
        \end{scope}

        \begin{scope}[xshift=11cm, local bounding box=linear_chain]
            \node[spin] (1b) at (0,0) {1};
            \node[spin] (2b) at (3,0) {2};
            \node[spin] (3b) at (6,0) {3};
            \draw[coupling] (1b) -- node[above, label] {$J_{12}=1$} (2b);
            \draw[coupling] (2b) -- node[above, label] {$J_{23}=1$} (3b);
            
            \node[label, align=center] at (3,-1.5) {
                $E_Q = -1$, \quad $E_{D_2} = 0$ \\
                $E_{D_2} - E_Q = 1$
            };
            
            \node[systemlabel] at (3.5,1) { 3 Spin Linear Chain};
        \end{scope}

        \begin{scope}[xshift=3cm, yshift=-5cm,local bounding box=triangle_frustrated]
            \node[spin] (1c) at (0,0) {1};
            \node[spin] (2c) at (-1.5,-2.6) {2};
            \node[spin] (3c) at (1.5,-2.6) {3};
            \draw[coupling] (1c) -- node[left, label] {$J_{12}=1$} (2c);
            \draw[coupling] (2c) -- node[below, label] {$J_{23}=1$} (3c);
            \draw[coupling] (3c) -- node[right, label] {$J_{13}=1$} (1c);
            
            \node[label, align=center, text width=4cm] at (0,-5) {
            $J_{12}=J_{23}=J_{13}=1$ \\
                $E_Q = -1.5$\\$E_{D_2}=E_{D_1}=1.5$ \\
                $E_{D_2} - E_Q = 3.0$
            };
            
            \node[systemlabel] at (0,0.5) {3 Spin Frustrated Triangle};
        \end{scope}

        \begin{scope}[yshift=-5cm, xshift=15cm, local bounding box=triangle_unfrustrated]
            \node[spin] (1d) at (0,0) {1};
            \node[spin] (2d) at (-1.5,-2.6) {2};
            \node[spin] (3d) at (1.5,-2.6) {3};
            \draw[coupling] (1d) -- node[left, label] {$J_{12}=1$} (2d);
            \draw[coupling] (2d) -- node[below, label] {$J_{23}=1$} (3d);
            \draw[coupling, very thick] (3d) -- node[right, label] {$J_{13}=2$} (1d);
            
            \node[label, align=center, text width=4cm] at (0,-5) {
                $J_{12}=J_{23}=1$, $J_{13}=2$ \\
                $E_Q = -2$, $E_{D_1} = 1$,$E_{D_2} = 3$ \\
                $E_{D_2} - E_Q = 5.0$\\$E_{D_1} - E_Q = 3.0$
            };
            
            \node[systemlabel] at (-0.5,0.5) {3 Spin Non-Frustrated Triangle};
        \end{scope}

        \begin{scope}[yshift=-14cm, xshift=7cm, local bounding box=asymmetric]
           \node[spin] (1b) at (0,0) {1};
            \node[spin] (2b) at (3,0) {2};
            \node[spin] (3b) at (6,0) {3};
            \draw[coupling] (1b) -- node[above, label] {$J_{12}=1$} (2b);
            \draw[coupling] (2b) -- node[above, label] {$J_{23}=1.1$} (3b);
            
            \node[label, align=center] at (3,-1.5) {$J_{12}=1,J_{23}=1.1,J_{13}=0$\\ 
                $E_{D_1} - E_Q = 3.15$
            };
            
            \node[systemlabel] at (2.5,1.0) { 3 Spin Asymmetric Linear Chain};
        \end{scope}

        \begin{pgfonlayer}{background}
            \node[box, fit=(two_spin) (linear_chain) (triangle_frustrated) 
                (triangle_unfrustrated) (asymmetric)] {};
        \end{pgfonlayer}
    \end{tikzpicture}
    \caption{Schematic representation of various spin systems and their resulting simplified excitation energies. Here, $J_{ij}$ is coupling strength between spins $i$ and $j$. $E_T$ , $E_S$ , $E_Q$, and $E_{D_i}$ represent triplet, singlet, quartet, and doublet energies,
respectively.
}
    \label{fig:spin_systems}
\end{figure*}
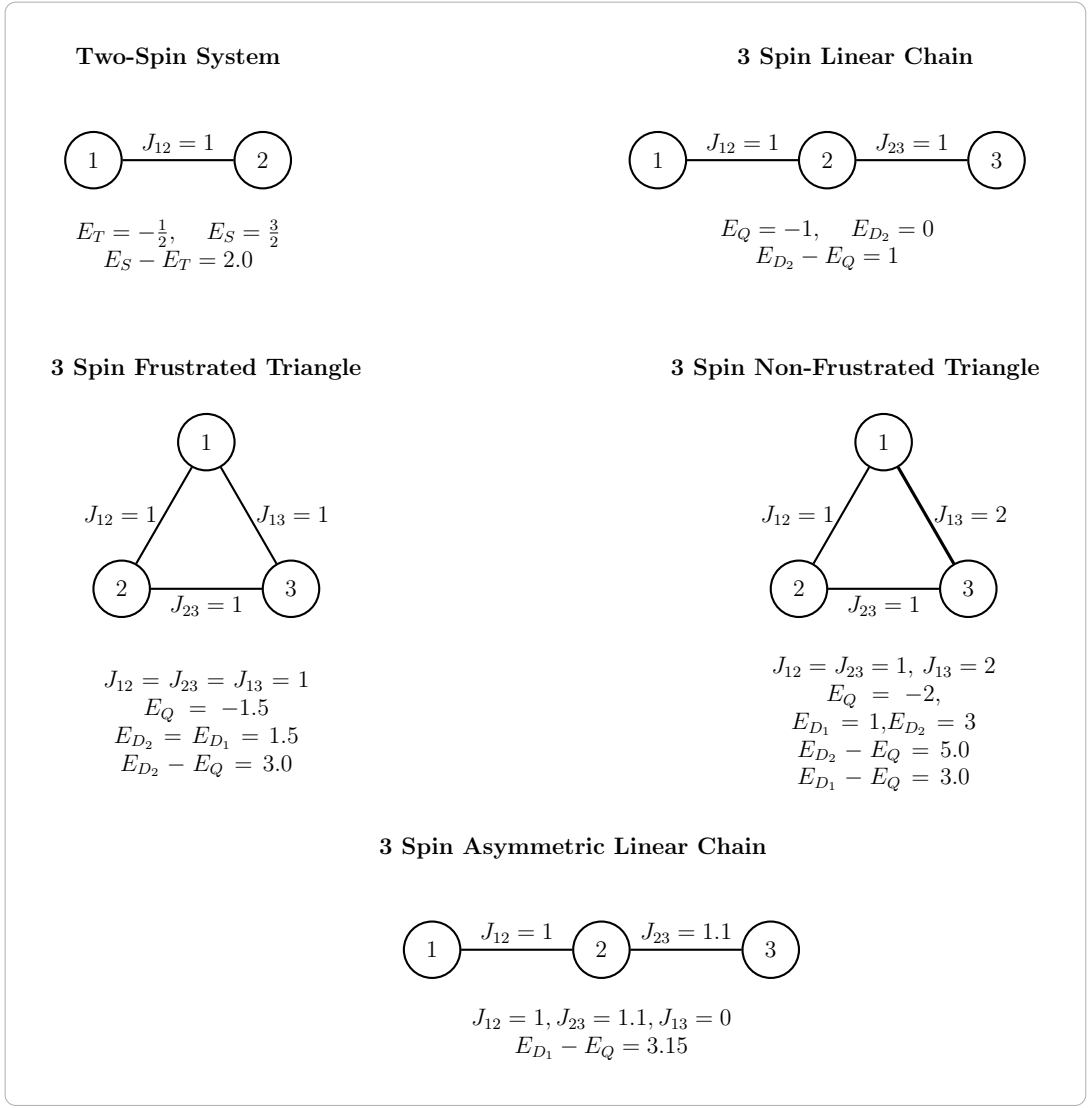

In this study, we focused on the spin-state energy gap of two and three-spin systems described by the Heisenberg Hamiltonian as illustrated in Fig.~\ref{fig:spin_systems}, and demonstrated a QPDE-based direct calculation of the energy gap using IBM Quantum processors. Through the multi-step pipeline optimization, the quantum circuit for the time evolution operator becomes a constant depth independent of the Trotter step, which allows us to compute the spin state energy gap with $>87$\% accuracy.

The paper is organized as follows: Section II presents the Heisenberg spin model and the QPDE algorithm. Section III investigates excitation energies in small spin systems with different geometries, considering both symmetric and asymmetric spin interactions using the QPDE algorithm. Finally, Section IV summarizes our findings.

\section{Model and Method}

\subsection{Heisenberg Spin Model}

The Heisenberg spin model \cite{Dirac1935,Heisenberg1928,EC1932} is a key theoretical approach for investigating quantum magnetism and the behavior of spin systems. This model effectively describes interactions between  spins, incorporating both ferromagnetic and antiferromagnetic phenomena through exchange coupling constants. For systems consisting of spin-\(\frac{1}{2}\) particles, their  dynamics are governed by the following Hamiltonian:

\begin{equation}
    H_{\text{H}} = -2\sum_{\langle i,j \rangle} J_{ij} \mathbf{S}_i \cdot \mathbf{S}_j 
\end{equation}

Here, $\mathbf{S}_i$ and $\mathbf{S}_j$ are total spin angular momentum operators, \(J_{ij}\) represents the exchange coupling constant between spins $i$ and $j$ , while \(\langle i,j \rangle\) denotes the summation over spin pairs.

The sign of \(J_{ij}\) determines the nature of the spin--spin interactions:

\begin{itemize}
    \item A positive \(J_{ij}\) indicates ferromagnetic behavior, where spins tend to align parallel to each other.
    \item A negative \(J_{ij}\) reflects antiferromagnetic interactions, where spins prefer to align antiparallel.
\end{itemize}

These exchange constants influence the magnetic ordering and the low-energy spectrum of the system. Systems with symmetric topologies and uniform couplings tend to have simpler energy spectra, while systems with asymmetric coupling result in more complex spectral structures.

We have the relation:
\begin{equation}
[H_{\text{H}}, \mathbf{S}^2] = 0    
\end{equation}
where the total spin angular momentum operator is given by:
\begin{equation}
    \mathbf{S}^2 = \mathbf{S}_x^2 + \mathbf{S}_y^2 + \mathbf{S}_z^2
\end{equation}
with $\mathbf{S}_x$, $\mathbf{S}_y$, and $\mathbf{S}_z$ representing the components of the spin angular momentum operator. From this commutation relation, we can use eigenfunctions of the $\mathbf{S}^2$ operators (hereafter referred to as spin eigenfunctions) as the basis of the wave function expansion of the Heisenberg Hamiltonian. Spin eigenfunctions can be constructed using the recurrence formula derived from the addition theorem of angular momentum~\cite{Pauncz2000,Sakurai2020,Sugisaki2016}, and those of two and three-spin systems are provided in Supplementary Materials.

\subsection{Principle of Quantum  Phase Difference Estimation (QPDE) Algorithm}

\begin{figure}[h]
  \includegraphics[width=0.5\textwidth]{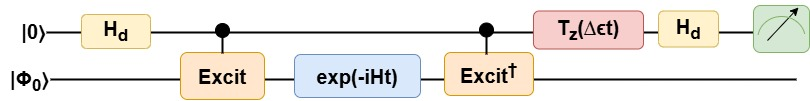}
  
    \caption{QPDE Quantum Circuit} 
    \label{fig:QPDE_circuit}
\end{figure}

A typical quantum circuit for the QPDE algorithm \cite{Sugisaki2021} is illustrated in Fig.\ref{fig:QPDE_circuit}. At first, a Hadamard ($\mathrm{H_d}$) gate is  applied to the ancillary qubit, creating a superposition of the \( |0\rangle \) and \( |1\rangle \) states. A controlled-Excit gate then generates a superposition of the approximate ground state (\( \Phi_0 \)) and excited state (\( \Phi_1 \)) wave functions:
\begin{equation}
  |\Phi_0\rangle = \sum _{j} c_j |\Psi_j \rangle  
\end{equation}
\begin{equation}
    Excit|\Phi_0\rangle = |\Phi_1\rangle = \sum _{k} d_k |\Psi_k \rangle
\end{equation}
Here, \{$\Psi$\} are eigenfunctions of the Hamiltonian, and $c_j$ and $d_k$ denote corresponding expansion coefficients. Subsequently, the control-free time evolution operator, the controlled-Excit$^\dagger$, the phase rotation gate $T_z(\Delta \epsilon t)$  defined as
\[
T_z(\Delta \epsilon t) =
\begin{pmatrix}
1 & 0 \\
0 & e^{i\Delta \epsilon t}
\end{pmatrix},
\]
and another Hadamard gate is applied. The ancilla qubit is then measured, and the energy gap is determined by computing the probability of obtaining the \( |0\rangle \) state, which is a function of the energy difference and the phase shift introduced during time evolution. Specifically, the probability is given by:

\begin{equation}
  P(0) = \frac{1}{2} \left( 1 + \sum _{j,k} |c_j|^2 |d_k|^2 \cos\left\{(\Delta E_{jk} - \Delta \epsilon)t \right\} \right)  
\end{equation}

In this equation, \(P(0)\) is the probability of measuring the state \(|0\rangle\),  \(\Delta E_{jk}\) represents the energy difference between $j$-th and $k$-th eigenstates, \(\Delta \epsilon\) is the phase shift applied during time evolution, and \(t\) is the time evolution duration. From this equation, if the approximated wave functions have a sufficiently large overlap with the eigenfunction of the corresponding target states, \( P(0) \) reaches its maximum around the point where \( \Delta \varepsilon \) equals the energy difference between the two targeted states. Thus, the energy gap can be determined by identifying the value of \( \Delta \epsilon \) that maximizes \( P(0) \). In QPDE, the quantum inference theorem is utilized to iteratively update the phase shift angle and refine the estimate of the energy gap, making it both more efficient and practical for implementation in quantum devices.

\subsection{Method of Computation}
The approximated ground state wave function (\(|\Phi_0\rangle \)) and controlled-Excit operator can be prepared using shallow-depth quantum circuits. The time evolution of these wave functions is implemented through the following steps:

First, the Heisenberg spin Hamiltonian is mapped to a qubit Hamiltonian by encoding the degrees of freedom of a spin into a qubit: $|\alpha\rangle \leftrightarrow |0\rangle$ and $|\beta\rangle \leftrightarrow |1\rangle$.  Then, the quantum circuit for the time evolution operator is constructed using the first-order Trotter--Suzuki decomposition \cite{Trotter1959,Suzuki1976,Hatano2005}. Notably, the time evolution operator of the Heisenberg Hamiltonian exhibits match-gate \cite{Valiant2002,BassmanOftelie2022} like characteristics, allowing for constant-depth quantum circuits even for extended evolution times or higher Trotter steps. This inherent structure makes it particularly well-suited for implementation on NISQ  hardware.

In the QPDE algorithm, the energy difference $\Delta \epsilon$ is calculated through the following procedure:

\begin{enumerate}
    \item The process begins with setting an initial estimate of the range where the actual energy difference $\Delta E$ locates, using a Gaussian function with the mean $\mu_{\mathrm{ini}}$ and standard deviation $\sigma_{\mathrm{ini}}$ that satisfy the relationship  $\mu_{\mathrm{ini}} - \sigma_{\mathrm{ini}} \le \Delta E \le \mu_{\mathrm{ini}} + \sigma_{\mathrm{ini}}$,  or a uniform distribution in the range between $\mu_{\mathrm{ini}} -\sigma_{\mathrm{ini}}$ and $\mu_{\mathrm{ini}} +\sigma_{\mathrm{ini}}$. 
    
    \item Run the quantum circuit shown in Fig.~\ref{fig:QPDE_circuit} with the setting $\mu_{\mathrm{ini}} - \sigma_{\mathrm{ini}} \le \Delta \epsilon \le \mu_{\mathrm{ini}} + \sigma_{\mathrm{ini}}$ and generate a $\Delta \varepsilon$ vs. $P(0)$ plot. The plot is fitted with a Gaussian function, and the range estimate is updated by multiplying two (initial and fitted) Gaussians.
    

    \item If the mean of the fitted Gaussian function $\mu_\mathrm{fit}$  falls outside the range \( (\mu_{\text{ini}} - \lambda \sigma_{\text{ini}}, \mu_{\text{ini}} + \lambda \sigma_{\text{ini}}) \), where $\lambda$ is a predetermined threshold, an adaptive strategy is employed: return to the second step with $\mu_{\mathrm{fit}}$ as $\mu_{\mathrm{ini}}$, while maintaining the prior standard deviation. In this study we used $\lambda = 0.6$. This ensures reliable convergence to values consistent with the expected excitation energy.

    \item  Convergence check. If the standard deviation of the updated distribution $\sigma_{\mathrm{updated}}$  falls below the predetermined tolerance threshold $E_{\mathrm{thre}}$, then the algorithm returns the mean of the updated Gaussian function $\mu_{\mathrm{updated}}$ as the calculated energy gap. Otherwise, return to the second step and replace $\mu_{\mathrm{ini}}$ and $\sigma_{\mathrm{ini}}$ with $\mu_{\mathrm{updated}}$ and $\sigma_{\mathrm{updated}}$, respectively. In this work we used $E_{\mathrm{thre}} = 0.4$.

\end{enumerate}

As clearly seen from Eq. (6), accuracy of the calculated energy difference depends on the fidelity of the approximated wave functions $|c_j|^2$ and $|d_k|^2$. At first we focus on the  symmetric systems (two-spin and three-spin configurations with linear, triangular frustrated, and non-frustrated arrangements). In these systems the spin eigenfunctions constructed based on the addition theorem of angular momentum are simultaneous eigenfunctions of the Heisenberg Hamiltonian (see Section 2 of Supplementary Materials for details), and we can exclude the effect of imperfect state preparation on the calculated energy gap, by using spin eigenfunctions as $|\Phi_0\rangle$ and $|\Phi_1\rangle$. We used the Gaussian function with $\mu_{\mathrm{ini}} = 0$ and $\sigma_{\mathrm{ini}} = 10.0$ as the initial estimate. The evolution time length $t$ of the initial step was set to be 0.2, so that $P(0)$ has a single upward peak in the search area.  
However, in the asymmetric scenario, without prior knowledge of eigenstates and eigenvalues, we select one of the spin eigenfunctions that is expected to have the maximum overlap with the target state as $|\Phi_0\rangle$ and $|\Phi_1\rangle$, and use a uniform distribution as the initial estimate with $\mu_{\mathrm{ini}} = \langle\Phi_1|H|\Phi_1\rangle - \langle\Phi_0|H|\Phi_0\rangle$. 

From Eq.~(6), we observe that $P(0)$ exhibits a cosine dependence on the evolution time length $t$, where increasing $t$ results in higher-frequency oscillations. Within the QPDE algorithm, $t$ is dynamically adjusted during the iterative process to capture a complete half-cycle by refining the initial estimate of $\Delta\epsilon$ range. This refinement is accomplished by updating the initial parameters ($\mu_{\mathrm{ini}}$ and $\sigma_{\mathrm{ini}}$) as described in Step~4.

We also employed pipeline optimization \cite{kharkov2022arlinebenchmarksautomatedbenchmarking} subroutines using Qiskit \cite{Qiskit2024} and Pytket \cite{Sivarajah2020} to minimize quantum circuit depth. Advanced noise suppression techniques of Qiskit, including Pauli Twirling \cite{PhysRevA.94.052325} and Dynamical Decoupling \cite{PhysRevApplied.20.064027}, are incorporated to improve measurement quality. These methods effectively minimize quantum hardware noise and enhance the accuracy of results obtained from the quantum processor.
\section{Results and Discussion}

 \begin{table*}[b]
\caption{QPDE simulation results for various spin systems on IBM quantum devices. For each system, the table lists the ground and excited state wave functions $|\Phi_0\rangle$ and $|\Phi_1\rangle$, the evolution time $t$, the number of Trotter steps used, the initial parameters (mean and standard deviation), the updated parameters, and the exact energy difference $\Delta E_{\text{Exact}}$. Configurations marked with (R) indicate an adaptive restart.}
\label{tab:quantum_systems_updated}
\begin{ruledtabular}
\begin{tabular}{c|c|c|c|c|c|c|c|c}
Ground State & Excited State & Time (t) & Trotter Steps & $\mu_{\mathrm{ini}} $ & $\sigma_{\mathrm{ini}}$ & $\mu_{\mathrm{update}}$ & $\sigma_{\mathrm{update}} $ & $\Delta E_{\text{Exact}}$ \\
\hline
\multicolumn{9}{c}{Two-Spin System (\( J_{12} = 1 \))/\texttt{ibm\_kyoto}} \\
\hline
$|T\rangle$ & $|S\rangle$ & 0.2 & 1 & 0.00 & 10.00 & 0.96 & 3.92 &  \\
& & 0.4 & 1 & 0.96 & 3.92 & 1.38 & 1.63 &  \\
& & 0.8 & 1 & 1.38 & 1.63 & 1.60 & 0.77 & 2.00 \\
& & 2.4 & 1 & 1.60 & 0.77 & 1.74 & 0.28 &  \\
\hline
\multicolumn{9}{c}{3 Spin Linear Chain (\( J_{12} = J_{23} = 1 \))/\texttt{ibm\_sherbrooke}} \\
\hline
$|Q\rangle$ & $|D2\rangle$ & 0.2 & 30 & 0.00 &10.00 & 0.56 & 4.06 &  \\
& & 0.4 & 60 & 0.56 & 4.06 & 0.74 & 1.84 &  \\
& & 1.0 & 150 & 0.74 & 1.84 & 0.79 & 0.78 & 1.00 \\
& & 4.2 & 620 & 0.79 & 0.78 & 0.88 & 0.32 & \\
\hline
\multicolumn{9}{c}{3 Spin Frustrated Triangle ( \( J_{12} = J_{23} = J_{13} = 1 \))/\texttt{ibm\_kyiv}} \\
\hline
$|Q\rangle$ & $|D2\rangle$ & 0.2 & 30 & 0.00 & 10.00 & 1.16 & 4.13 &  \\
& & 0.4 & 60 & 1.16 & 4.13 & 1.76 & 1.75 &  \\
& & 0.8 & 120 & 1.76 & 1.75 & 2.06 & 0.72 & 3.00 \\
& & 0.8(R) & 120 & 2.06 & 1.75 & 2.34 & 0.78 &  \\
& & 1.6 & 240 & 2.34 & 0.78 & 2.51 & 0.37 & \\
& & 1.6(R) & 240 & 2.51 & 0.78 & 2.64 & 0.34 &  \\
\hline
\multicolumn{9}{c}{3 Spin Non-Frustrated Triangle (Doublet 1) (\( J_{12} = J_{23} = 1 \) while \( J_{13} = 2 \))/\texttt{ibm\_kyiv}} \\
\hline
$|Q\rangle$ & $|D1\rangle$ & 0.2 & 30 & 0.00 & 10.00 & 0.85 & 4.15 &  \\
& & 0.4 & 60 & 0.85 & 4.15 & 1.47 & 1.72 &  \\
& & 0.6 & 90 & 1.47 & 1.72 & 1.76 & 0.74 &  \\
& & 0.6(R) & 90 & 1.76 & 1.72 & 2.03 & 0.75 & 3.00\\
& & 0.6(R) & 90 & 2.03 & 1.72 & 2.28 & 0.86 & \\
& & 1.6 & 240 & 2.28 & 0.86 & 2.44 & 0.43 &  \\
& & 1.6(R) & 240 & 2.44 & 0.86 & 2.56 & 0.40 &  \\
\hline
\multicolumn{9}{c}{3 Spin Non-Frustrated Triangle (Doublet 2) (\( J_{12} = J_{23} = 1 \) while \( J_{13} = 2 \))/\texttt{ibm\_kyiv}} \\
\hline
$|Q\rangle$ & $|D2\rangle$ & 0.2 & 30 & 0.00 & 10.00 & 1.75 & 4.15 & \\
& & 0.4 & 60 & 1.75 & 4.15 & 2.77 & 2.01 &  \\
& & 0.4(R) & 60 & 2.77 & 4.15 & 3.48 & 1.75 &  \\
& & 0.6 & 90 & 3.48 & 1.75 & 3.83 & 0.77 & 5.00 \\
& & 0.6(R) & 90 & 3.83 & 1.75 & 4.11 & 0.78 &  \\
& & 0.6(R) & 90 & 4.11 & 1.75 & 4.36 & 0.77 &  \\
& & 1.8 & 270 & 4.36 & 0.77 & 4.55 & 0.40 &  \\
& & 1.8(R) & 270 & 4.55 & 0.77 & 4.67 & 0.31 &  \\
\hline
\multicolumn{9}{c}{3 Spin Asymmetric Linear Chain (\( J_{12} = 1 \), \( J_{23} = 1.1 \), and \( J_{13} = 0 \))/\texttt{ibm\_kyiv}} \\
\hline
$|Q\rangle$ & $ |D1\rangle$ & 1.2 & 180 & 3.00 & 1.15 & 2.89 & 1.02 &  \\
& & 1.8 & 300 & 2.89 & 1.02 & 3.03 & 0.43 & 3.15 \\
& & 4.2 & 620 & 3.03 & 0.43 & 3.04 & 0.18 &  \\
\end{tabular}
\end{ruledtabular}
\end{table*}

We implemented our QPDE circuit across different spin systems using 127 qubits IBM Quantum Eagle processors , strategically selecting the least busy hardware based on the lowest two-qubit error rates, average error per layer gate (EPLG) \cite{mckay2023benchmarkingquantumprocessorperformance}, and circuit layer operations per second (CLOPS) \cite{wack2021qualityspeedscalekey}. Implementing these systems required 3 qubits for two-spin and 4 qubits for three-spin configurations (including one ancilla qubit).  The details of the results of QPDE analysis for each configuration are given in table \ref{tab:quantum_systems_updated}. For all cases, classical results ($\Delta E_{Exact}$) were verified using Mathematica application \cite{Mathematica}, with additional configuration plots available in the Supplementary Materials.

\subsection{Two-spin systems}

\begin{figure*}[htbp]
\centering
\subfloat[$t = 0.2$\label{fig:2spin_a}]{%
\includegraphics[width=0.45\textwidth]{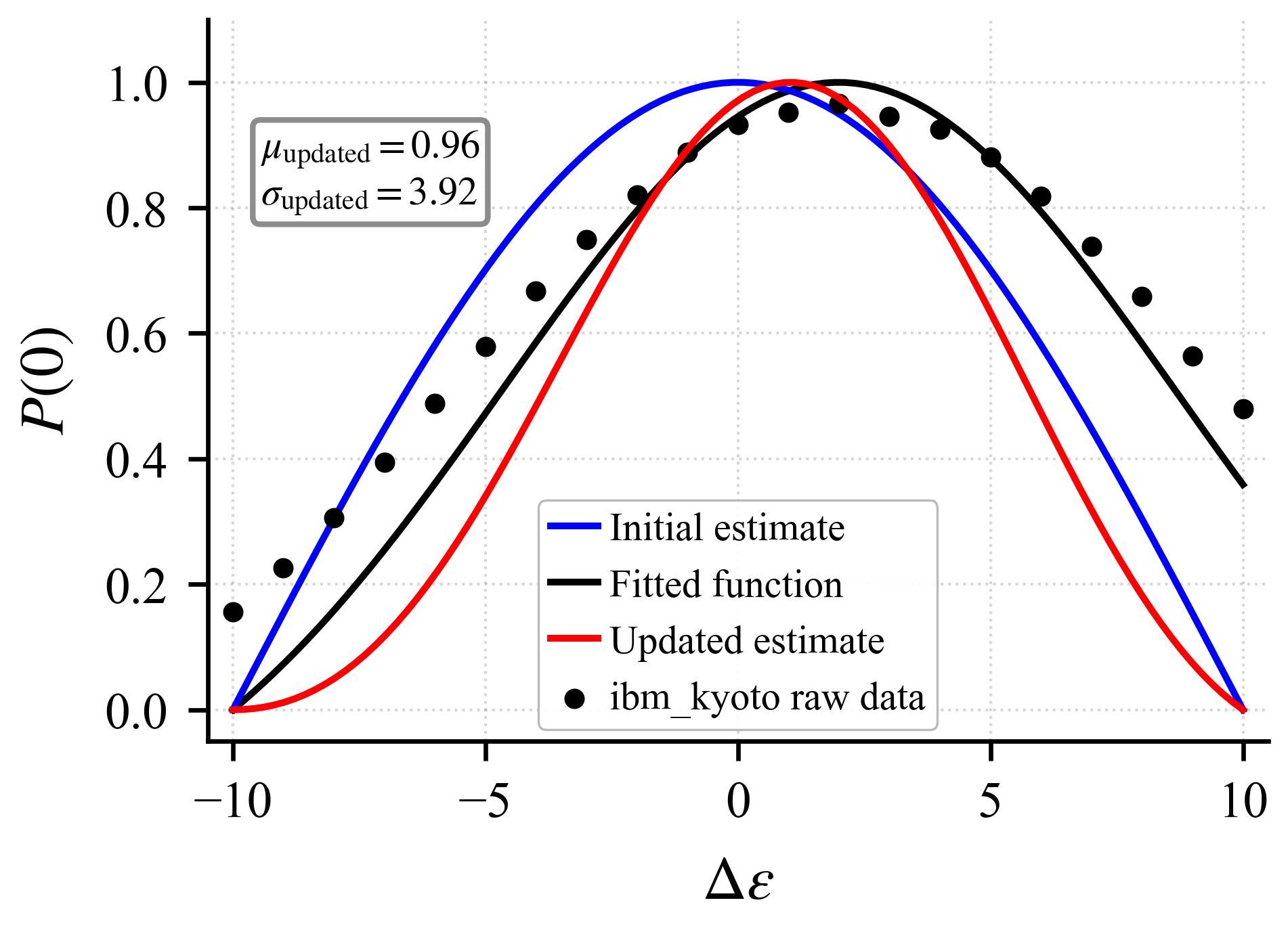}
}
\hfill
\subfloat[$t = 0.4$\label{fig:2spin_b}]{%
\includegraphics[width=0.45\textwidth]{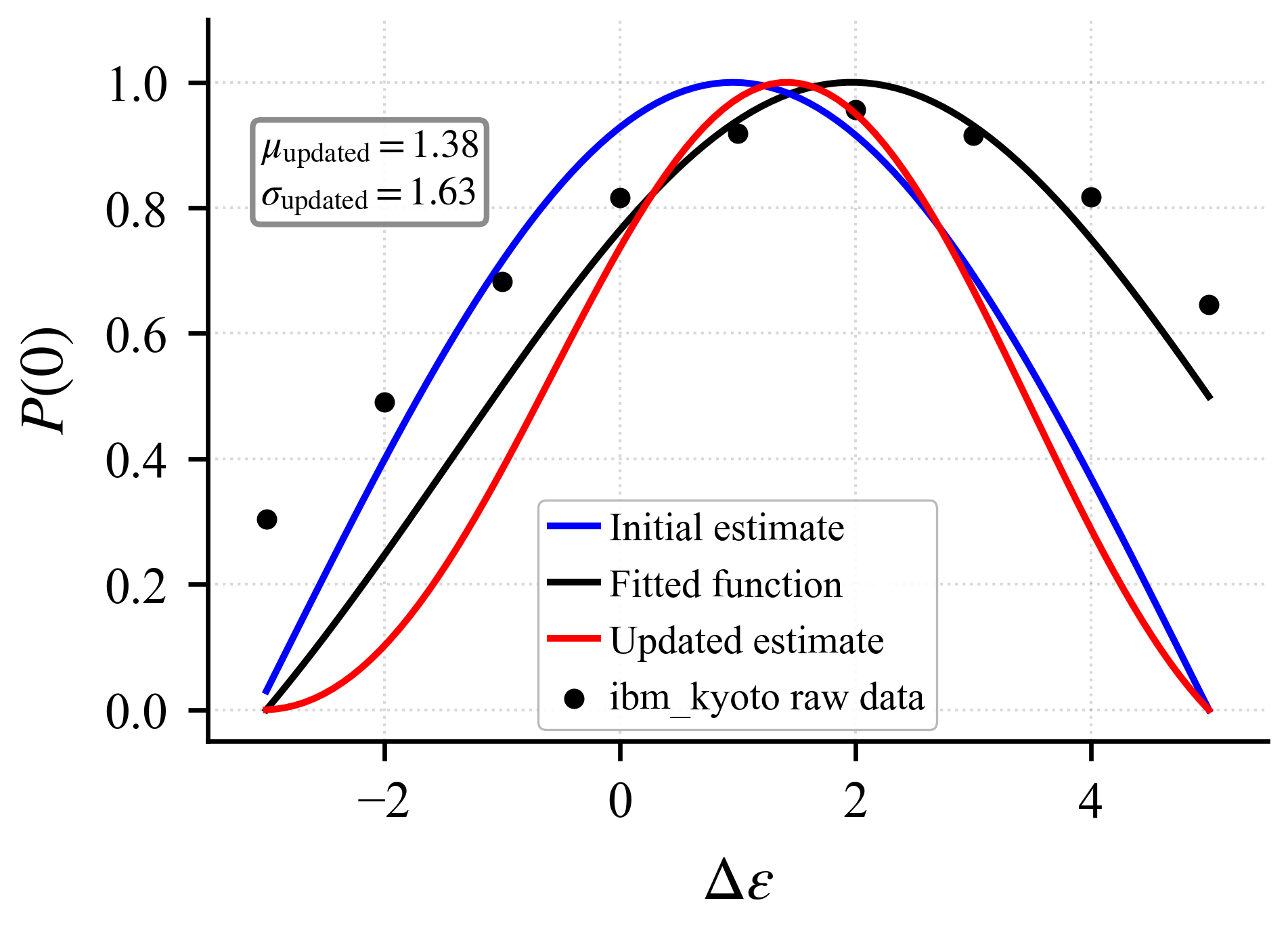}
}
\vspace{0.5cm}

\subfloat[$t = 0.8$\label{fig:2spin_c}]{%
    \includegraphics[width=0.45\textwidth]{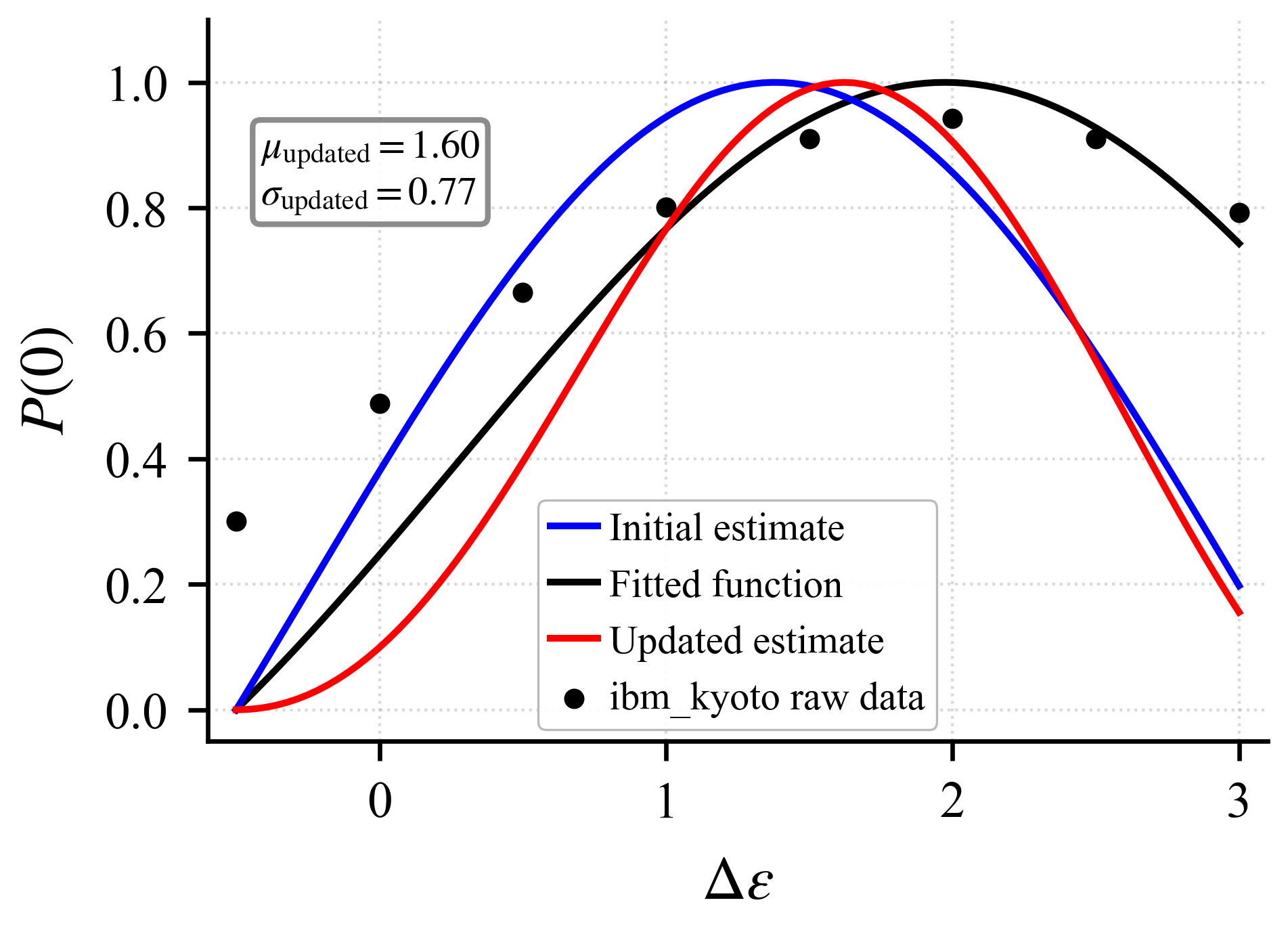}
}
\hfill
\subfloat[$t = 2.4$\label{fig:2spin_d}]{%
    \includegraphics[width=0.45\textwidth]{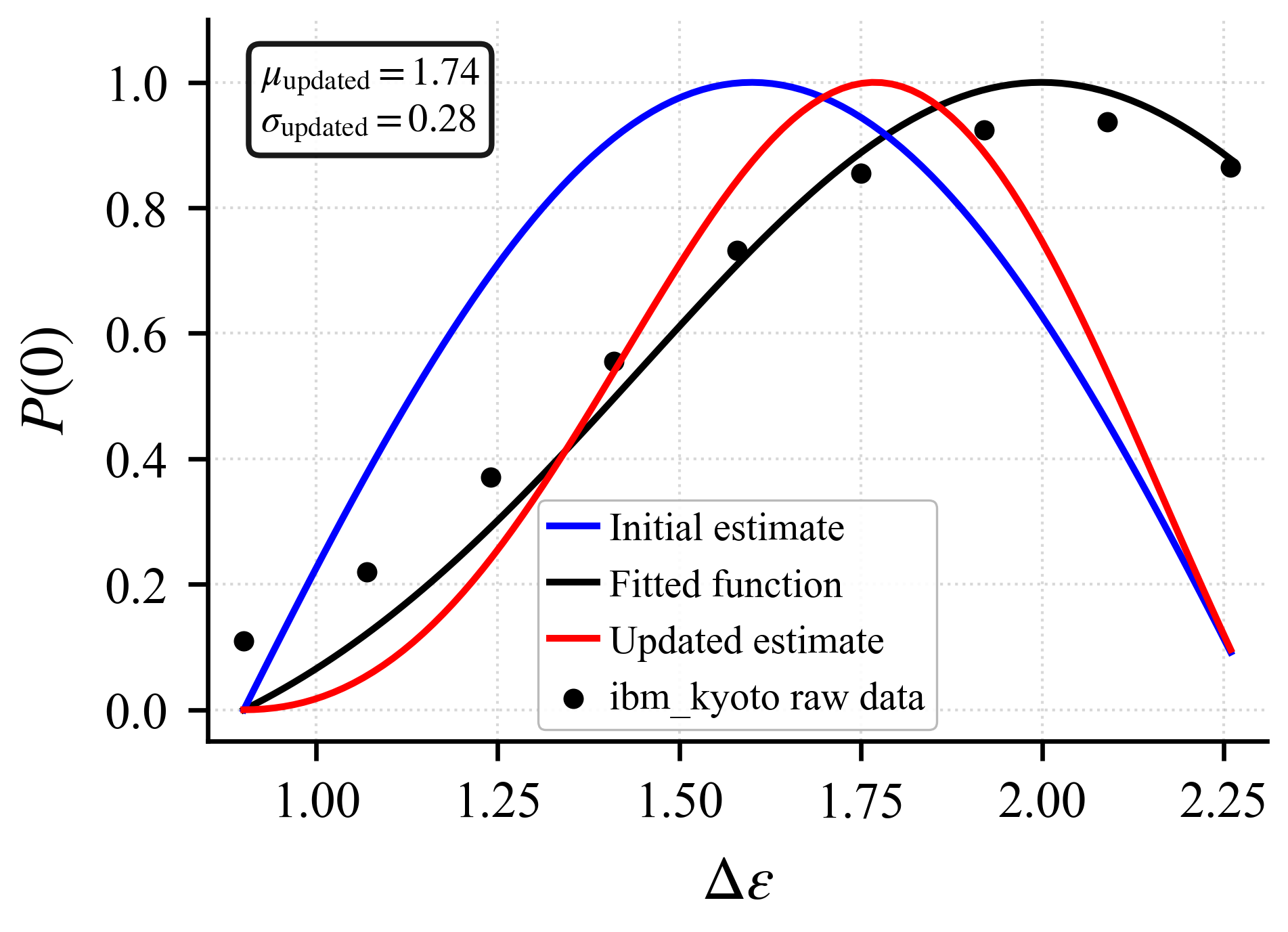}
}

\caption{ Initial estimate (blue), \texttt{ibm\_kyoto} raw data (black dots), fitted function (black), and updated estimate (red) distributions for the QPDE algorithm in a two-spin Heisenberg system at different evolution time lengths.
(a) $t = 0.2$: Updated mean $0.96 \pm 3.92$, 
(b) $t = 0.4$: Improved estimate $1.38 \pm 1.63$, 
(c) $t = 0.8$: Next updated mean $1.60 \pm 0.77$, 
(d) $t = 2.4$: Final convergence with updated mean $1.74 \pm 0.28$.}
\label{fig:updated_distributions_2spin}
\end{figure*}

We follow a similar strategy for each spin configuration to calculate the excitation energy using the QPDE algorithm. Figure~\ref{fig:updated_distributions_2spin}(a--d) demonstrates the execution on a real quantum device (\texttt{ibm\_kyoto}) for two-spin systems. With the interaction strength set to \( J_{12} = 1 \), the system was prepared in the ground and excited states \( |\Phi_0\rangle = |T\rangle = \frac{1}{\sqrt{2}} (|01\rangle + |10\rangle) \) and \( |\Phi_1\rangle = |S\rangle = \frac{1}{\sqrt{2}} (|01\rangle - |10\rangle) \), respectively, where $|T\rangle$ and $|S\rangle$ represent spin eigenfunctions of the triplet and singlet states, respectively. For the triplet state, we chose the $M_s = 0$ wave function, because controlled-Excit gate can be realized by a controlled-Z gate.
A single Trotter step implements the time evolution operator in the two-spin Hamiltonian with commuting terms. The circuit depth before and after transpilation in quantum hardware was 13 and 26, respectively, and six two-qubit gates were required to prepare the QPDE circuit.
In Figure~\ref{fig:updated_distributions_2spin}(a--d), three distribution plots are presented: initial estimate (blue), $ibm\_kyoto$ raw data (black dots)(obtained from quantum hardware with 5000 shots), fitted function (black) and updated estimate (red).
Iterations continue through Figure~\ref{fig:updated_distributions_2spin}(b--d) until the updated standard deviation falls below 0.4. The updated mean in the final iteration (\( t = 2.4 \)) was \(1.74 \pm 0.28\), achieving 87\% accuracy relative to the exact value of 2.0. 

\subsection{Three-spin systems}
Next, we explore various geometrical configurations in three-spin systems, including linear chains, triangular arrangements, and asymmetric chains, by systematically varying the exchange coupling constants. In these systems, the Hamiltonian terms generally do not commute, necessitating an increased number of Trotter steps to minimize Trotter errors for a given final evolution time. However,  the number of two-qubit gates and circuit depth grow linearly with the number of Trotter steps, and in the context of NISQ-era quantum hardware, additional gate errors are introduced. While Qiskit's optimization level 3 routing reduces the number of two-qubit gates and circuit depth, this reduction is insufficient to fully mitigate errors.

By leveraging the match gate-like structure of the Heisenberg Hamiltonian's time evolution operator, we apply a multi-step optimization: Qiskit optimization level 0 $\rightarrow$ Pytket optimization level 2 $\rightarrow$ Qiskit optimization level 3. This yields a constant-depth circuit, with two-qubit gate counts and depth invariant to Trotter steps. While IBM hardware's heavy-hex lattice  connectivity increases gate counts and depth in ISA (Instruction Set Architecture) circuits, the overall circuit complexity remains independent of Trotter steps, enabling efficient simulations of three-spin systems on near-term quantum hardware.

For example, simulating a three-spin linear chain at $t = 0.2$ with 30 Trotter steps results in an unoptimized depth of 368 and 186 two-qubit gates. After optimization, depth reduces to 54 and gates to 22. Post-mapping to IBM hardware, depth and gates stabilize at 184 and 46, respectively. This holds for longer time evolution: at $t = 4.2$ with 620 Trotter steps, unoptimized depth and gates reach 7448 and 3726, but after optimization, they remain constant at 54 and 22 (184 and 46 in ISA circuits).

Our optimization ensures circuit complexity does not scale with Trotter steps, enabling feasible simulations of longer time evolution on NISQ-era hardware without prohibitive errors. Detailed metrics for various configurations are provided in the Supplementary Material.

In the case of a three-spin linear configuration where only adjacent spins interact ($J_{12} = J_{23} \ne 0$ and $J_{13} = 0$), two non-degenerate doublets emerge. We set the interaction strengths as \( J_{12} = J_{23} = 1 \). In this case, as discussed in the Supplementary Materials, spin eigenfunctions are simultaneous eigenfunctions of the Heisenberg Hamiltonian, with the eigenenergies $E_Q = -1$, $E_{D_1} = 2$, and $E_{D_2} = 0$. The $M_s = S$ spin eigenfunctions of the spin-quartet ($Q$), spin-doublet 1 ($D_1$), and spin-doublet 2 ($D_2$) states are given as follows in the qubit basis.
\begin{eqnarray}
    |Q\rangle = |000\rangle
\end{eqnarray}
\begin{eqnarray}
    |D_1\rangle = \frac{1}{\sqrt{6}}(2|010\rangle - |100\rangle - |001\rangle)
\end{eqnarray}
\begin{eqnarray}
    |D_2\rangle = \frac{1}{\sqrt{2}}(|001\rangle - |100\rangle)
\end{eqnarray}

In the QPDE demonstration, we used $|\Phi_0\rangle = |Q\rangle$ and $|\Phi_1\rangle = |D_2\rangle$. 
Using four qubits, we implemented this system on \texttt{ibm\_sherbrooke}.
The final energy difference, as listed in Table~\ref{tab:quantum_systems_updated}, was \( 0.88 \pm 0.32 \), demonstrating an accuracy of 88\%  compared to the classical reference value of 1.0.

In a three-spin equilateral triangular arrangement with equal exchange couplings \( J_{12} = J_{23} = J_{13} \), the system exhibits geometric spin frustration in the spin-doublet state, where competing interactions prevent simultaneous energy minimization, resulting in a degenerate doublet state. We uniformly set interaction strengths to \( J_{12} = J_{23} = J_{13} = 1 \), which results in $E_Q = -1.5$ and $E_{D_1} 
= E_{D_2} = 1.5$. Consequently, we selected one of the degenerate doublet excited states to quantitatively analyze the quartet--doublet energy gap. In this study we used \texttt{ibm\_kyiv} processor, and setting $|\Phi_0\rangle = |Q\rangle$ and $|\Phi_1\rangle = |D_2\rangle$. We selected $|D_2\rangle$ state, because implementation of the controlled-Excit gate is simpler. 

At \( t = 0.8 \) and \( t = 1.6 \), the fitted mean, \( \mu_{\text{fit}} \), fell outside the range \( (\mu_{\text{ini}} - 0.6\sigma_{\text{ini}}, \mu_{\text{ini}} + 0.6\sigma_{\text{ini}}) \), requiring QPDE restarts with the following adjustments:

\begin{itemize}
    \item \( t = 0.8 \): $\mu_{\mathrm{ini}}$ = 2.06, $\sigma_{\mathrm{ini}}$ = 1.75
    \item \( t = 1.6 \): $\mu_{\mathrm{ini}}$ = 2.51, $\sigma_{\mathrm{ini}}$ = 0.78
\end{itemize}

The final excitation energy, as listed in Table~\ref{tab:quantum_systems_updated}, was \( 2.64 \pm 0.34 \), achieving an accuracy of 88\% relative to the classical value of 3.0.

When two couplings are equal and the third differs (\(J_{12} = J_{23} \neq J_{13}\)), the degeneracy of the doublets is lifted, while spin eigenfunctions remain the simultaneous eigenfunctions of the Heisenberg Hamiltonian. 
We modified the previous configuration by setting \( J_{12} = J_{23} = 1 \) while \( J_{13} = 2 \) ($E_Q = -2$, $E_{D_1} = 1$, and $E_{D_2} = 3$) and implemented this configuration also on \texttt{ibm\_kyiv} processor. First we calculated the energy gap between the quartet and doublet 1 state, by using $|\Phi_0\rangle = |Q\rangle$ and $|\Phi_1\rangle = |D_1\rangle$. In this case also restart conditions  were required at \( t = 0.6 \) and \( t = 1.6 \), with initial means iteratively refined:

\begin{itemize}
    \item \( t = 0.6 \): 
    \begin{itemize}
        \item First restart: $\mu_{\mathrm{ini}}$ = 1.76, $\sigma_{\mathrm{ini}}$ = 1.72
        \item Second restart: $\mu_{\mathrm{ini}}$ = 2.03, $\sigma_{\mathrm{ini}}$ = 1.72
    \end{itemize}
    \item \( t = 1.6 \): $\mu_{\mathrm{ini}}$ = 2.44, $\sigma_{\mathrm{ini}}$ = 0.86
\end{itemize}

The final energy difference was \(2.56 \pm 0.40\), demonstrating (given in table \ref{tab:quantum_systems_updated})  88\% accuracy.

The same configuration was examined on the same quantum hardware, focusing on the doublet 2 state as the excited state. 

In this case, multiple QPDE restarts were necessary:

\begin{itemize}
    \item \( t = 0.4 \): $\mu_{\mathrm{ini}}$ = 2.77, $\sigma_{\mathrm{ini}}$ = 4.15
    \item \( t = 0.6 \): 
    \begin{itemize}
        \item First restart: $\mu_{\mathrm{ini}}$ = 3.83, $\sigma_{\mathrm{ini}}$ = 1.75
        \item Second restart: $\mu_{\mathrm{ini}}$ = 4.11, $\sigma_{\mathrm{ini}}$ = 1.75
    \end{itemize}
    \item \( t = 1.8 \): $\mu_{\mathrm{ini}}$ = 4.55, $\sigma_{\mathrm{ini}}$ = 0.77
\end{itemize}

The final energy difference was \( 4.67 \pm 0.31 \), with 93.4\% accuracy relative to the exact value of 5.0. 

Lastly, we investigated a unique case of asymmetric exchange coupling strength on \texttt{ibm\_kyiv} processor with \( J_{12} = 1 \), \( J_{23} = 1.1 \), and \( J_{13} = 0 \), where the spin eigenfunctions are no longer simultaneous eigenfunctions of the Heisenberg Hamiltonian. The spin eigenfunction of the quartet state is still an eigenfunction of the Heisenberg Hamiltonian, but the spin-doublet eigenfunctions of the Heisenberg Hamiltonian is written as a linear combination of $|D_1\rangle$ and $|D_2\rangle$, because $\langle D_1|H_\mathrm{H}|D_2\rangle = \sqrt{3}(J_{12} - J_{23})/2 \ne 0$. The approximate ground and excited states are respectively $|\Phi_0\rangle = |Q\rangle$ and $|\Phi_1\rangle = |D_1\rangle$. The final energy difference was determined to be \(3.04 \pm 0.18\), in good agreement with the true energy gap $\Delta E_{Exact} = 3.15$.

We observed that the singlet--triplet energy gap (two-spin system) and doublet--quartet energy gap (three-spin system) exhibits minimal deviation from the exact energy difference ($\Delta E_{Exact}$). This slight discrepancy is attributed to hardware noise and trotter decomposition error. These marginal deviations demonstrate the QPDE algorithm's capability to predict energy gaps accurately. The precision of results also depends on the updated estimate threshold (0.4 in this study), with more stringent thresholds potentially improving accuracy.

\section{Conclusion}
In this work, we have demonstrated QPDE-based direct calculation of the spin-state energy gap of two and three-spin Heisenberg Hamiltonians, achieving accuracy rates ranging from 85.33\% to 93\% using real superconducting quantum hardware. 
Thanks to the match gate-like structure of the time evolution operator of the Heisenberg Hamiltonian, the pipeline optimization using Pytket and Qiskit gives a constant depth quantum circuit for the time evolution operator, enabling efficient implementation of longer time evolutions.
The systematic refinement of updated estimates, coupled with our noise suppression techniques, establishes a reliable framework for excitation energy calculations in the presence of hardware noise and experimental constraints.
Our comprehensive analysis demonstrates the remarkable versatility and robustness of the QPDE algorithm across various spin system configurations. These findings not only validate the theoretical framework but also establish practical guidelines for implementing quantum simulations on current NISQ devices, paving the way for more complex quantum many-body simulations in the future.



\section{Acknowledgments}
 B.P. thanks Dr. Ranjan Modak (IIT Tirupati) for his continuous support throughout this project. K.S. acknowledges support from Quantum Leap Flagship Program (Grant No. JPMXS0120319794) from Ministry of Education, Culture, Sports, Science and Technology (MEXT), Japan, Center of Innovations for Sustainable Quantum AI (JPMJPF2221) from Japan Science and Technology Agency (JST), Japan, and Grants-in-Aid for Scientific Research C (21K03407) and for Transformative Research Area B (23H03819) from Japan Society for the Promotion of Science (JSPS), Japan. K.S. thanks Naoki Yamamoto, Hiroshi Yano, Shu Kanno, Kosuke Ito, and Kaito Wada, for useful discussions.

\bibliographystyle{apsrev4-2}
\bibliography{main.bib}

\end{document}


\title{Supplementary
 Materials}
\date{}
\maketitle

\tableofcontents 
\newpage 

\section{Spin Eigenfunctions}

Eigenfunctions of the spin ${\mathbf S}^2$ operator (spin eigenfunctions) are typically expressed as \(X(N,S,M_s;d)\), where:
\begin{itemize}
    \item \(N\) is the number of spins,
    \item \(S\) is the total spin quantum number,
    \item \(M_s\) is the magnetic spin quantum number, and 
    \item $d$ is the index to distinguish degenerate spin states,
\end{itemize}
and spin eigenfunctions can be constructed using the following recurrence formula derived from addition theorem of angular momentum. 
\begin{eqnarray*}
    X\left(N, S+\frac{1}{2}, M+\frac{1}{2}\right) &=& \frac{\sqrt{S+M+1}X(N-1, S, M)X(1, \frac{1}{2}, \frac{1}{2})}{\sqrt{2S+1}} \notag \\
    &+& \frac{\sqrt{S-M}X(N-1, S, M+1)X(1, \frac{1}{2}, -\frac{1}{2})}{\sqrt{2S+1}} \notag
\end{eqnarray*}
\begin{eqnarray*}
    X\left(N, S-\frac{1}{2}, M+\frac{1}{2}\right) &=& \frac{-\sqrt{S-M}X(N-1, S, M)X(1, \frac{1}{2}, \frac{1}{2})}{\sqrt{2S+1}} \notag \\
    &+& \frac{\sqrt{S+M+1}X(N-1, S, M+1)X(1, \frac{1}{2}, -\frac{1}{2})}{\sqrt{2S+1}} \notag
\end{eqnarray*}
\begin{eqnarray*}
    X\left(N, S+\frac{1}{2}, M-\frac{1}{2}\right) &=& \frac{\sqrt{S-M+1}X(N-1, S, M)X(1, \frac{1}{2}, -\frac{1}{2})}{\sqrt{2S+1}} \notag \\
    &+& \frac{\sqrt{S+M}X(N-1, S, M-1)X(1, \frac{1}{2}, \frac{1}{2})}{\sqrt{2S+1}} \notag
\end{eqnarray*}
\begin{eqnarray*}
    X\left(N, S-\frac{1}{2}, M-\frac{1}{2}\right) &=& \frac{-\sqrt{S+M}X(N-1, S, M)X(1, \frac{1}{2}, -\frac{1}{2})}{\sqrt{2S+1}} \notag \\
    &+& \frac{\sqrt{S-M+1}X(N-1, S, M-1)X(1, \frac{1}{2}, \frac{1}{2})}{\sqrt{2S+1}} \notag
\end{eqnarray*}

The spin eigenfunctions of the two-spin systems are given as follows:

\textbf{Spin-triplet states:}

\begin{eqnarray*}
X(2,1,1) &=& |\alpha\alpha\rangle \notag \\
X(2,1,0) &=& \frac{1}{\sqrt{2}}(|\alpha\beta\rangle+|\beta\alpha\rangle) \notag \\
X(2,1,-1) &=& |\beta\beta\rangle
\end{eqnarray*}

\textbf{Spin-singlet state:}

\begin{eqnarray}
X(2,0,0) = \frac{1}{\sqrt{2}}(|\alpha\beta\rangle-|\beta\alpha\rangle) \notag
\end{eqnarray}

In the case of three-spin systems, spin eigenfunctions include:

\textbf{Spin-quartet state:}

\begin{eqnarray*}
X\left(3,\frac{3}{2},\frac{3}{2}\right) &=& |\alpha\alpha\alpha\rangle \notag \\
X\left(3,\frac{3}{2},\frac{1}{2}\right) &=& \frac{1}{\sqrt{3}}(|\alpha\alpha\beta\rangle+|\alpha\beta\alpha\rangle+|\beta\alpha\alpha\rangle) \notag \\
X\left(3,\frac{3}{2},-\frac{1}{2}\right) &=& \frac{1}{\sqrt{3}}(|\beta\beta\alpha\rangle+|\beta\alpha\beta\rangle+|\alpha\beta\beta\rangle) \notag \\
X\left(3,\frac{3}{2},-\frac{3}{2}\right) &=& |\beta\beta\beta\rangle \notag \\
\end{eqnarray*}

\textbf{Spin-doublet state 1:}

\begin{eqnarray*}
X\left(3,\frac{1}{2},\frac{1}{2};1\right) &=& \frac{1}{\sqrt{6}}(2|\alpha\alpha\beta\rangle-|\alpha\beta\alpha\rangle-|\beta\alpha\alpha\rangle) \notag \\
X\left(3,\frac{1}{2},-\frac{1}{2};1\right) &=& \frac{1}{\sqrt{6}}(2|\beta\beta\alpha\rangle-|\alpha\beta\beta\rangle-|\beta\alpha\beta\rangle) \notag
\end{eqnarray*}

\textbf{Spin-doublet state 2:}

\begin{eqnarray*}
X\left(3,\frac{1}{2},\frac{1}{2};2\right) &=& \frac{1}{\sqrt{2}}(|\alpha\beta\alpha\rangle-|\beta\alpha\alpha\rangle) \notag \\
X\left(3,\frac{1}{2},-\frac{1}{2};2\right) &=& \frac{1}{\sqrt{2}}(|\alpha\beta\beta\rangle-|\beta\alpha\beta\rangle) \notag
\end{eqnarray*}

\newpage
\section{Matrix Representation of the Heisenberg Hamiltonian}

For a two-spin system, the Heisenberg Hamiltonian in matrix form, expressed in the spin configuration basis\{$|\alpha\alpha\rangle
,|\alpha\beta\rangle
,|\beta\alpha\rangle
,|\beta\beta\rangle
$\} is given below.
\[
H_\mathrm{H} =
\begin{bmatrix}
-J_{12}/2 & 0 & 0 & 0 \\
0 & J_{12}/2 & -J_{12} & 0 \\
0 & -J_{12} & J_{12}/2 & 0\\
0 & 0 & 0 & -J_{12}/2
\end{bmatrix}
\]
Here $J_{12}$ represents the exchange coupling constant between spins ${\mathbf S}_1$ and ${\mathbf S}_2$.

When the Heisenberg Hamiltonian matrix is rewritten in the spin eigenfunction basis ($X(2,1,1)$, $X(2, 1, 0)$, $X(2, 1, -1)$ and $X(2, 0, 0)$), the following matrix can be obtained, and the singlet--triplet energy gap is calculated to be $\Delta E_{\mathrm{ST}} = E_\mathrm{S} - E_\mathrm{T} = 2J_{12}$.

\[
H_\mathrm{H;\ se} =
\begin{bmatrix}
-J_{12}/2 & 0 & 0 & 0 \\
0 & -J_{12}/2 & 0 & 0 \\
0 & 0 & -J_{12}/2 & 0\\
0 & 0 & 0 & 3J_{12}/2
\end{bmatrix}
\]
\\\\
For a three-spin system, the Hamiltonian also can be represented similarly in the spin configuration basis 
\{$|\alpha\alpha\alpha\rangle
,|\alpha\alpha\beta\rangle
,|\alpha\beta\alpha\rangle
,|\beta\alpha\alpha\rangle
,|\alpha\beta\beta\rangle
,|\beta\alpha\beta\rangle
,|\beta\beta\alpha\rangle
,|\beta\beta\beta\rangle
$\}. 

\[
H_\mathrm{H} =
\begin{bmatrix}
A_{11} & 0 & 0 & 0 & 0 & 0 & 0 & 0 \\
0 & A_{22} & A_{23} & A_{24} & 0 & 0 & 0 & 0 \\
0 & A_{23} & A_{33} & A_{34} & 0 & 0 & 0 & 0 \\
0 & A_{24} & A_{34} & A_{44} & 0 & 0 & 0 & 0 \\
0 & 0 & 0 & 0 & A_{55} & A_{56} & A_{57} & 0 \\
0 & 0 & 0 & 0 & A_{56} & A_{66} & A_{67} & 0 \\
0 & 0 & 0 & 0 & A_{57} & A_{67} & A_{77} & 0 \\
0 & 0 & 0 & 0 & 0 & 0 & 0 & A_{88}
\end{bmatrix}
\]
where $A_{ij}$ is given as follows. 
\begin{eqnarray*}
A_{11} = A_{88} &=& -\frac{J_{12} + J_{23} + J_{13}}{2}, \notag \\
A_{22} = A_{77} &=& \frac{-J_{12} + J_{23} + J_{13}}{2}, \notag \\
A_{33} = A_{66} &=& \frac{J_{12} + J_{23} - J_{13}}{2}, \notag \\
A_{44} = A_{55} &=& \frac{J_{12} - J_{23} + J_{13}}{2}, \notag \\
A_{23} = A_{67} &=& -J_{23}, \notag \\
A_{24} = A_{57} &=& -J_{13}, \notag \\
A_{34} = A_{56} &=& -J_{12} \notag
\end{eqnarray*}

Here, $J_{ij}$ represent the exchange coupling constant between spins ${\mathbf S}_i$ and ${\mathbf S}_j$ respectively. When the Heisenberg Hamiltonian is rewritten in the spin eigenfunction basis, we can obtain the following matrix.

\[
H_\mathrm{H;\ se} =
\begin{bmatrix}
B_{QQ} & 0 & 0 & 0 & 0 & 0 & 0 & 0 \\
0 & B_{QQ} & 0 & 0 & 0 & 0 & 0 & 0 \\
0 & 0 & B_{QQ} & 0 & 0 & 0 & 0 & 0 \\
0 & 0 & 0 & B_{QQ} & 0 & 0 & 0 & 0 \\
0 & 0 & 0 & 0 & B_{D_1D_1} & 0 & B_{D_1D_2} & 0 \\
0 & 0 & 0 & 0 & 0 & B_{D_1D_1} & 0 & B_{D_1D_2} \\
0 & 0 & 0 & 0 & B_{D_1D_2} & 0 & B_{D_2D_2} & 0 \\
0 & 0 & 0 & 0 & 0 & B_{D_1D_1} & 0 & B_{D_2D_2} \\
\end{bmatrix}, 
\]
where
\begin{eqnarray*}
B_{QQ} &=& -\frac{1}{2}(J_{12} + J_{23} + J_{13}), \notag \\
B_{D_1D_1} &=& \frac{1}{2}(-J_{12} + 2J_{23} + 2J_{13}), \notag \\
B_{D_2D_2} &=& \frac{3}{2}J_{12}, \notag \\
B_{D_1D_2} &=& \frac{\sqrt{3}}{2}(J_{13} - J_{23}). \notag
\end{eqnarray*}

From this, for three-spin systems, spin eigenfunctions of the spin-doublet states become simultaneous eigenfunctions of the Heisenberg Hamiltonian, when $J_{13} = J_{23}$.

\newpage

\section{Results Of Different Spin Configurations}

\begin{figure*}[htbp]
\centering
\subfloat[$t = 0.2$\label{fig:linear_chain_a}]{%
\includegraphics[width=0.4\textwidth]{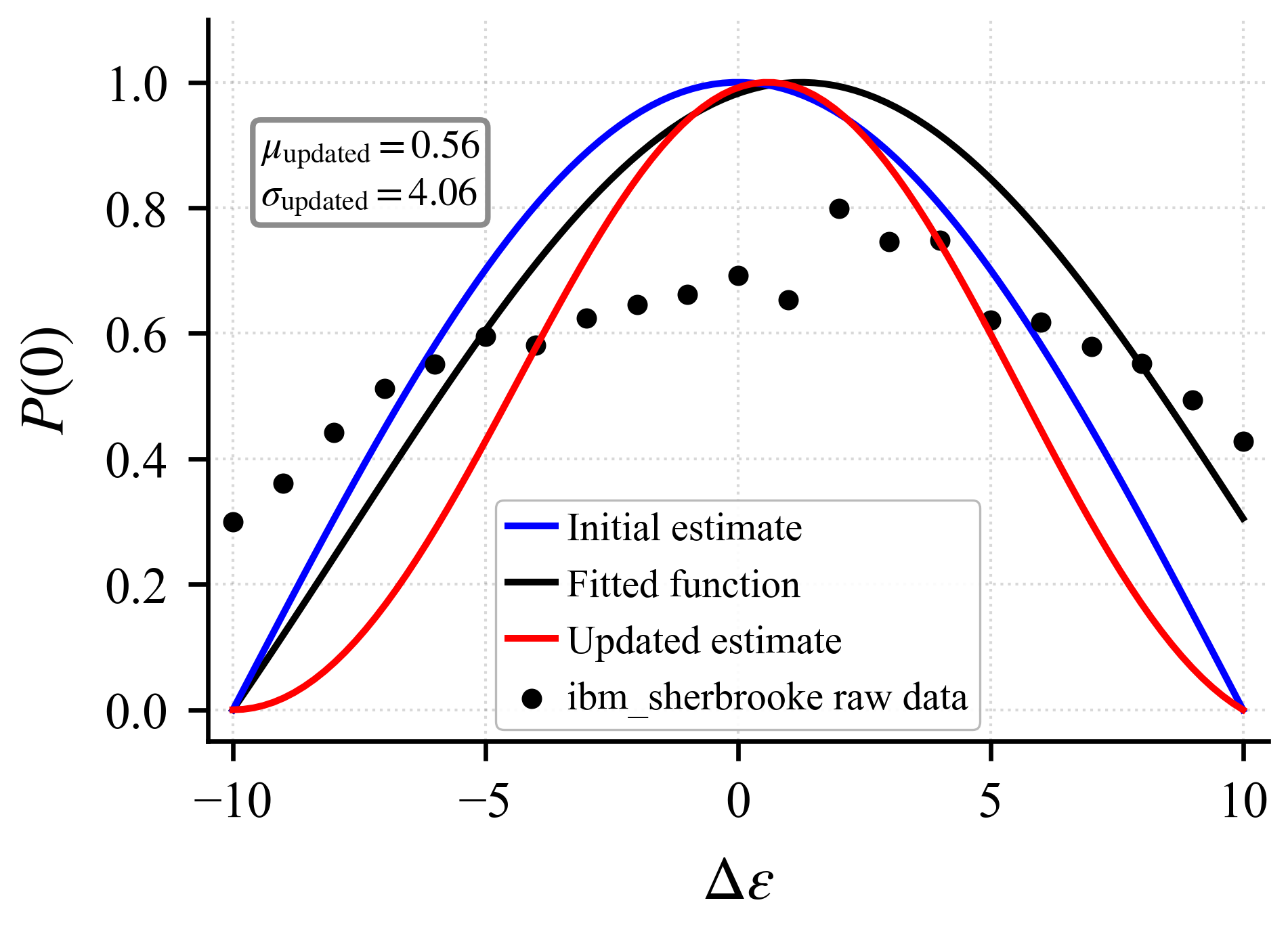}
}
\hfill
\subfloat[$t = 0.4$\label{fig:linear_chain_b}]{%
\includegraphics[width=0.4\textwidth]{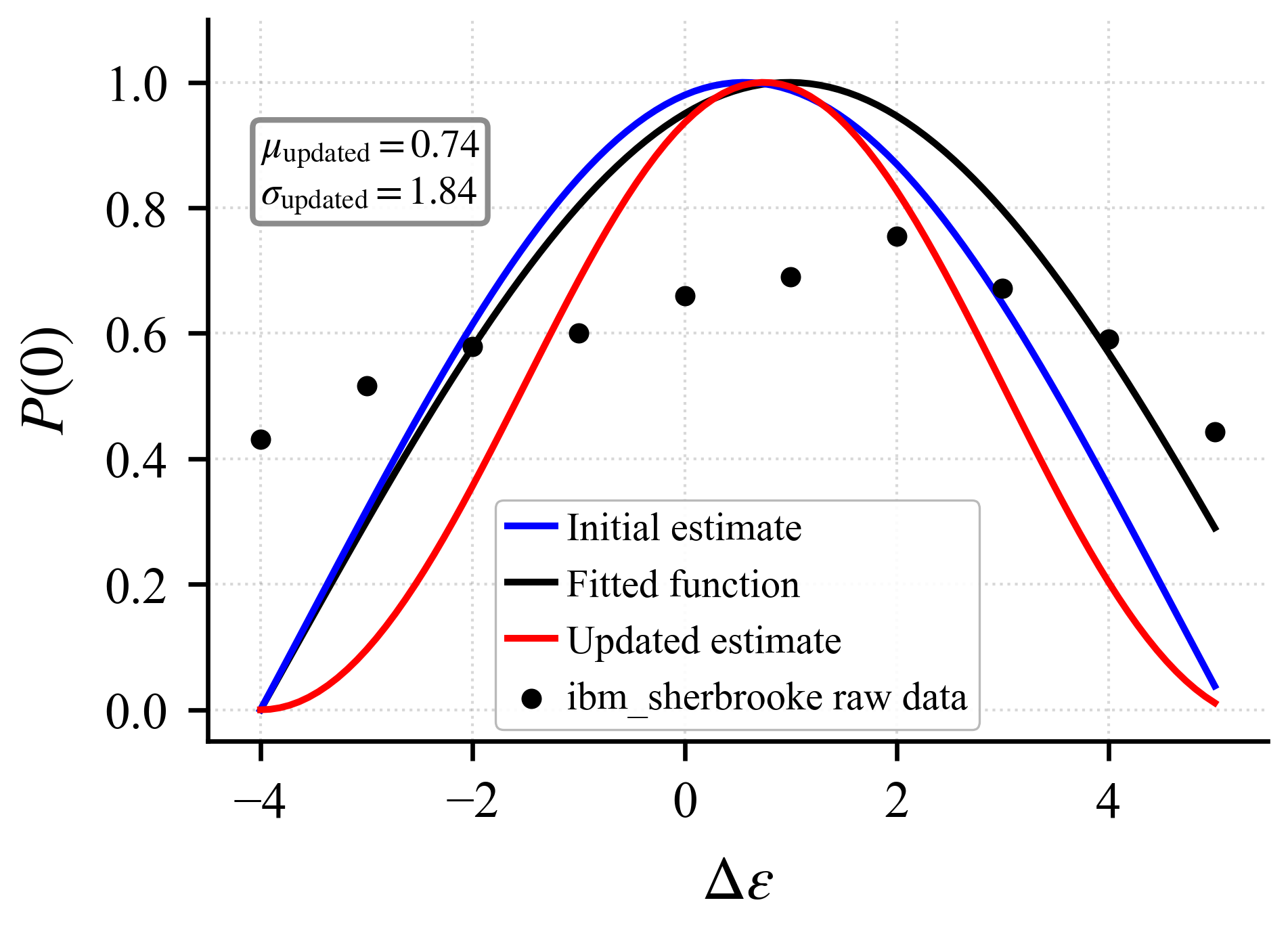}
}
\vspace{0.5cm}

\subfloat[$t = 1.0$\label{fig:linear_chain_c}]{%
    \includegraphics[width=0.4\textwidth]{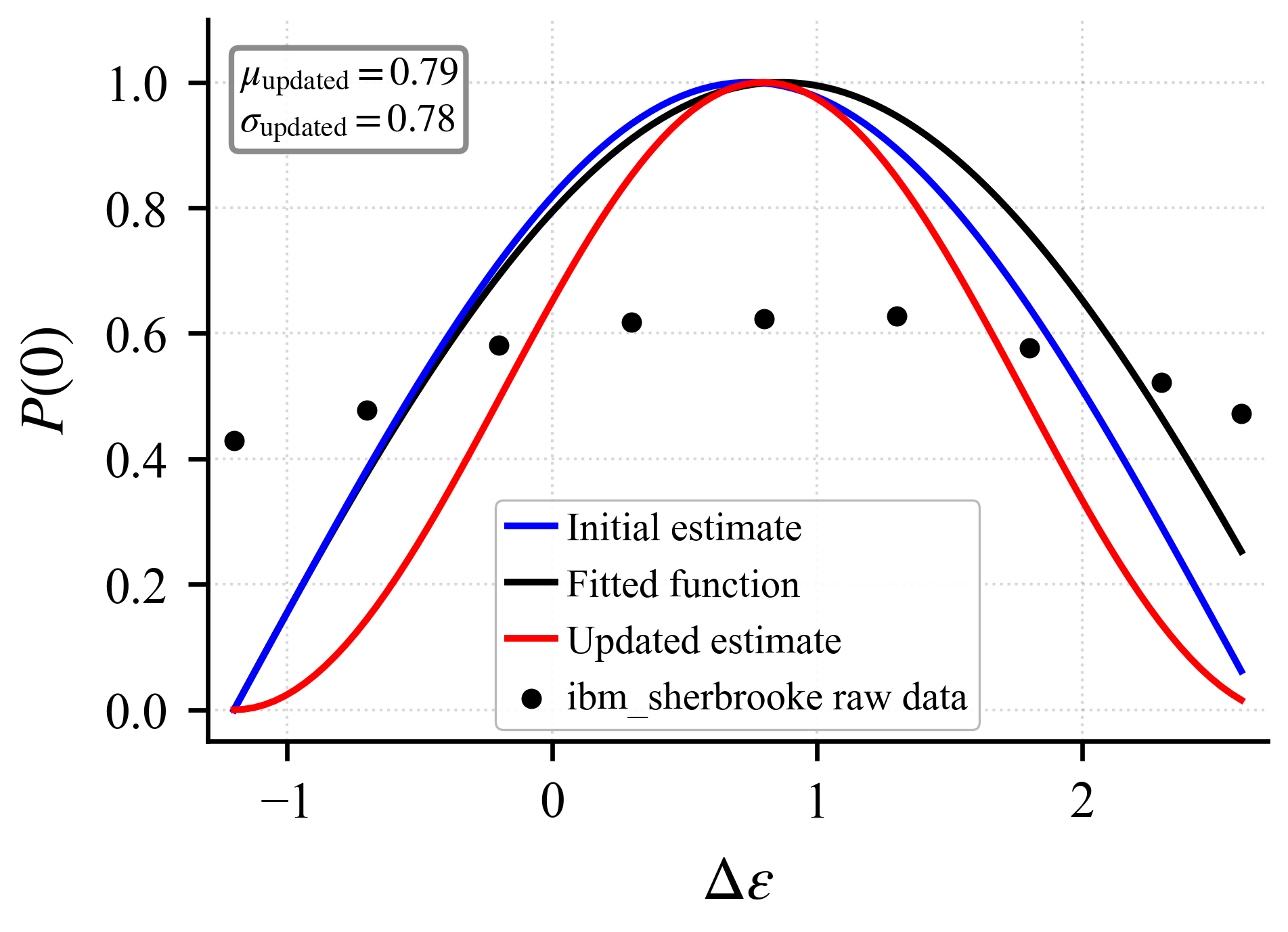}
}
\hfill
\subfloat[$t = 4.2$\label{fig:linear_chain_d}]{%
    \includegraphics[width=0.4\textwidth]{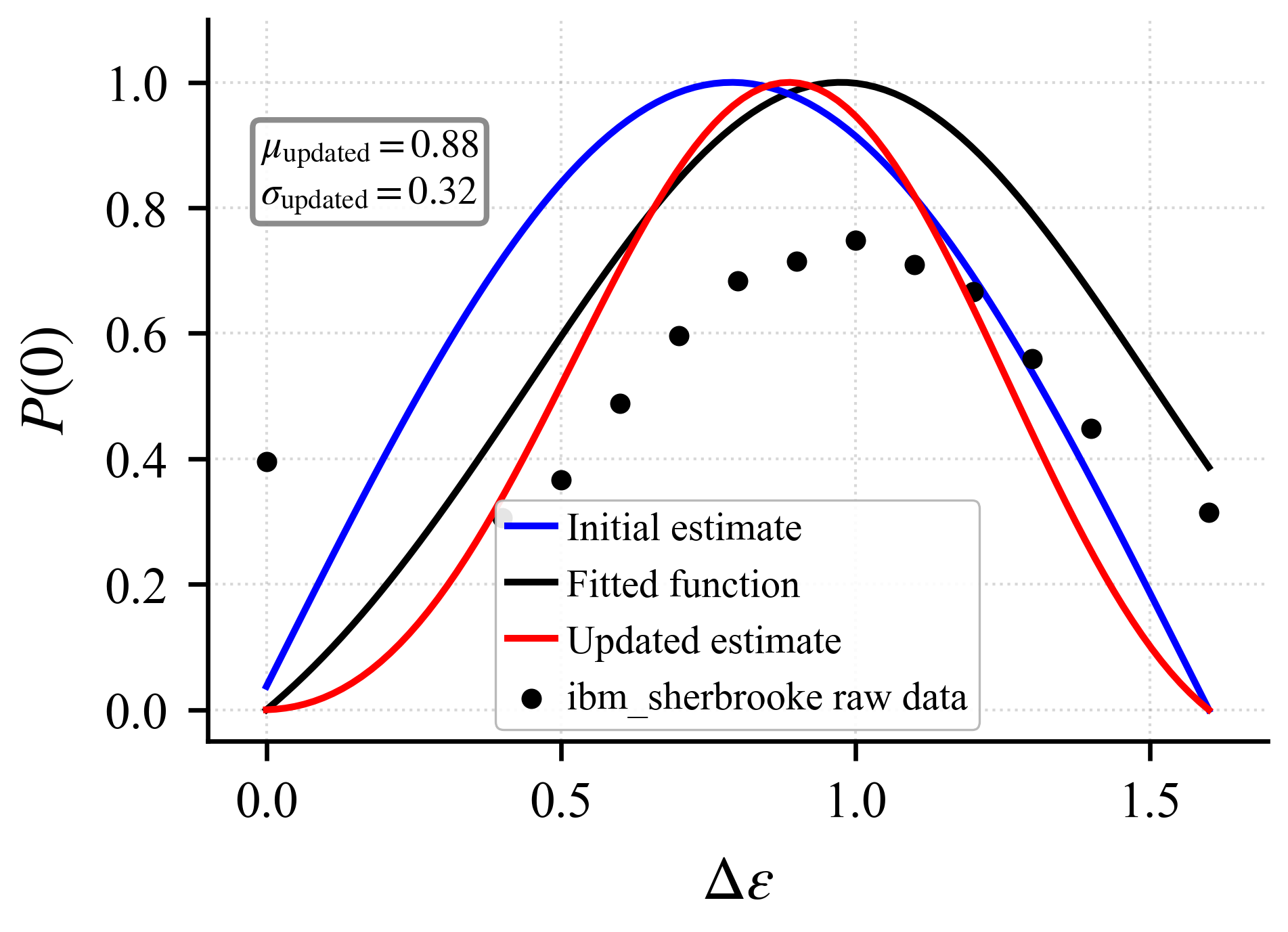}
}

\caption{The plots show the initial estimate (blue), \texttt{ibm\_sherbrooke} raw data (black dots), fitted function (black), and updated estimate (red) distributions across different evolution times for three-spin linear chain. (a) $t = 0.2$:  updated mean $0.56 \pm 4.06$ (b) $t = 0.4$:  updated mean $0.74 \pm 1.84$ (c) $t = 1.0$: updated mean $0.79 \pm 0.78$ (d) $t = 4.2$: updated mean $0.88 \pm 0.32$.}
\label{fig:updated_distributions_3spin_linear_chain}
\end{figure*}

\begin{figure*}[htbp]
\centering
\subfloat[$t = 0.2$\label{fig:triangular_frustration_a}]{%
\includegraphics[width=0.45\textwidth]{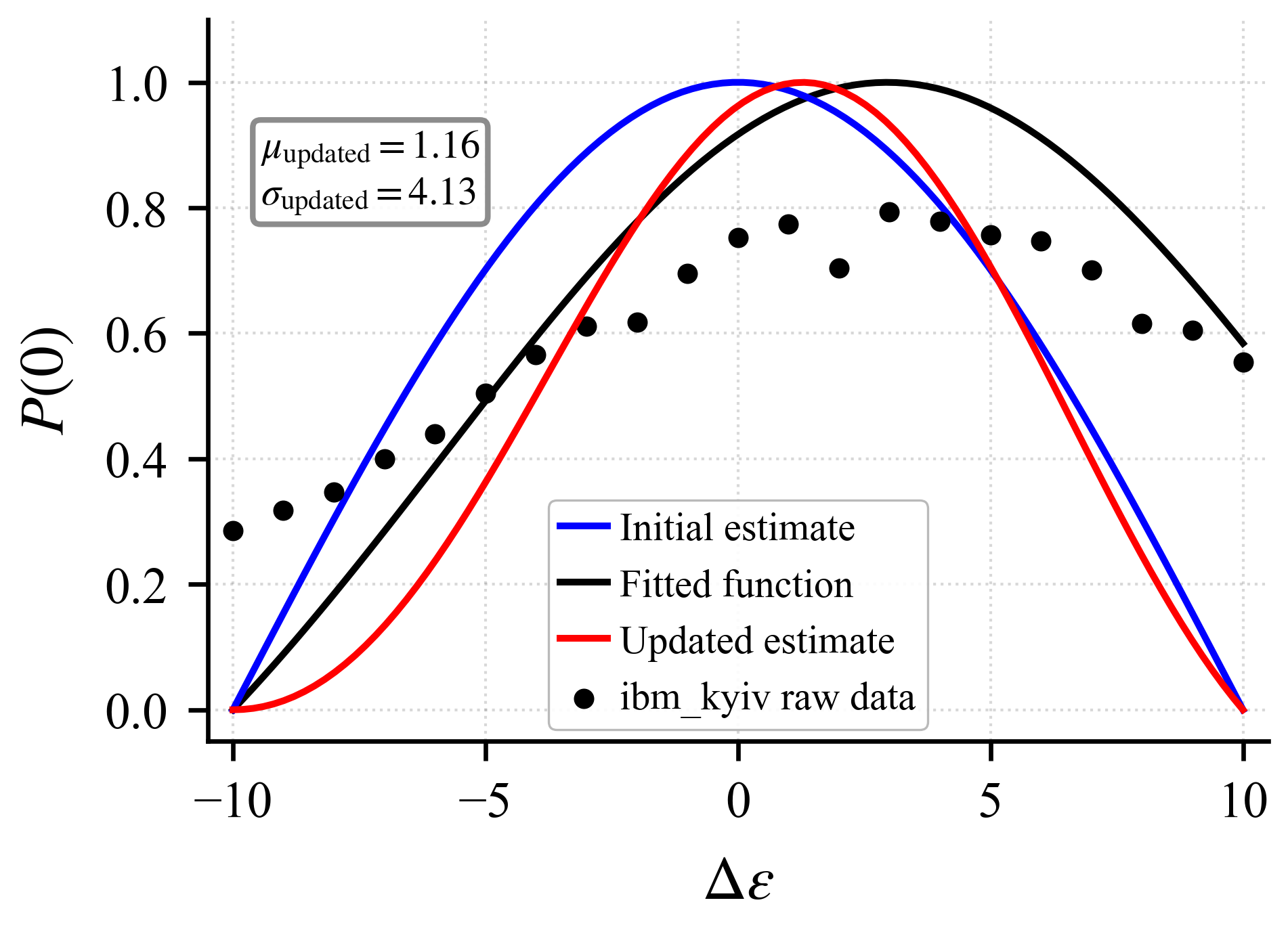}
}
\hfill
\subfloat[$t = 0.4$\label{fig:triangular_frustration_b}]{%
\includegraphics[width=0.45\textwidth]{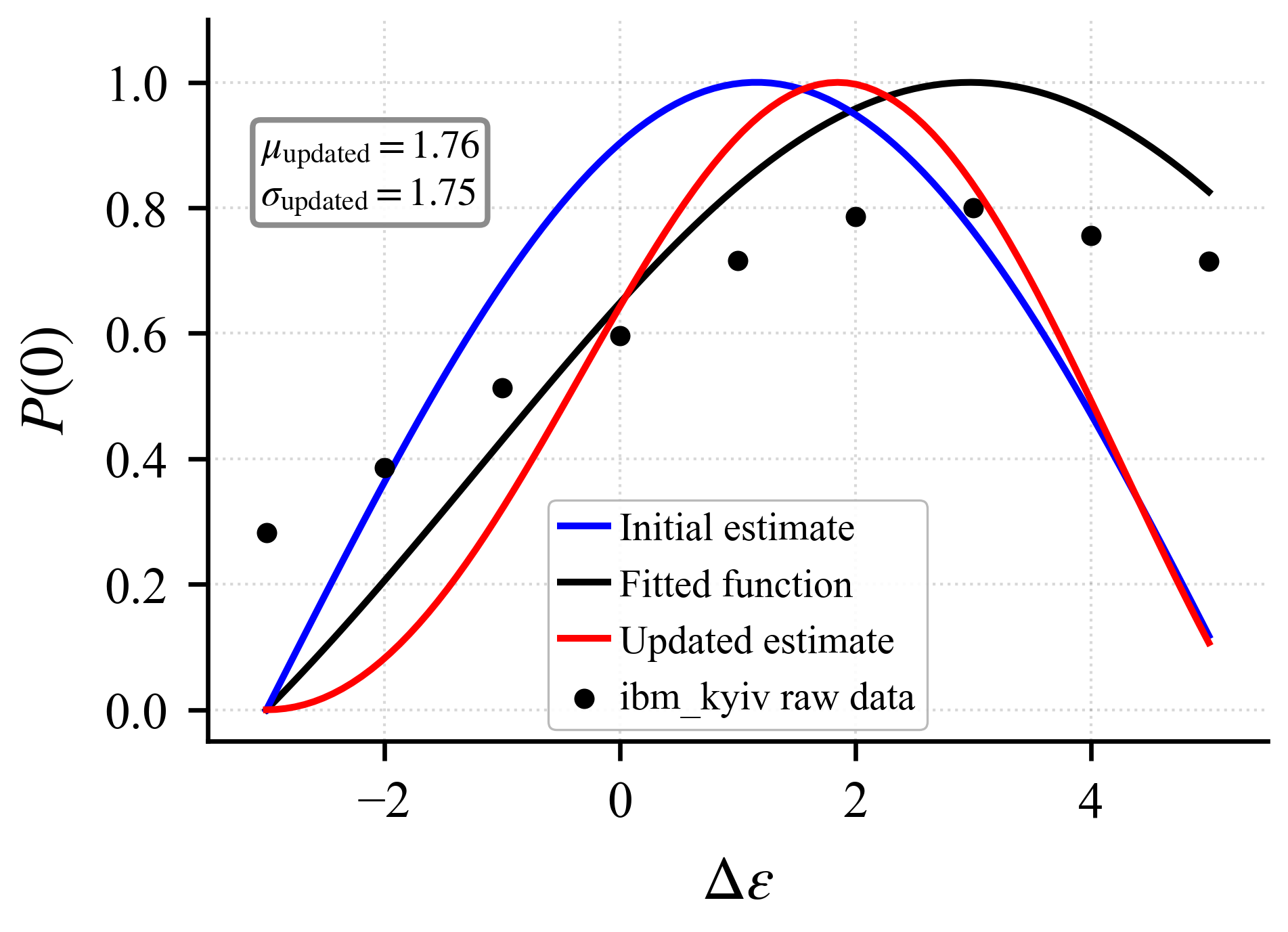}
}
\vspace{0.5cm}

\subfloat[$t = 0.8$\label{fig:triangular_frustration_c}]{%
    \includegraphics[width=0.45\textwidth]{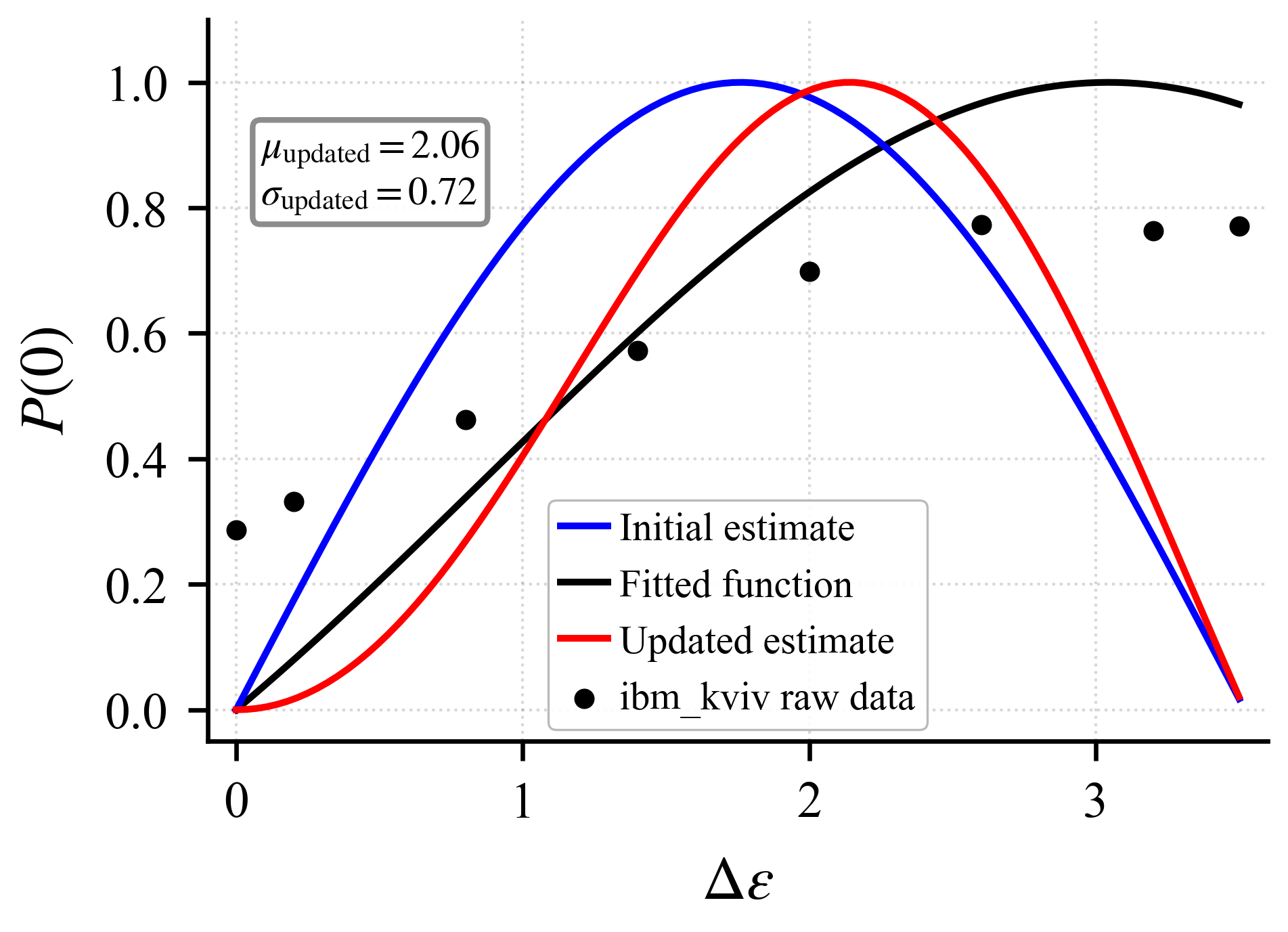}
}
\hfill
\subfloat[$t = 0.8$\label{fig:triangular_frustration_d}]{%
    \includegraphics[width=0.45\textwidth]{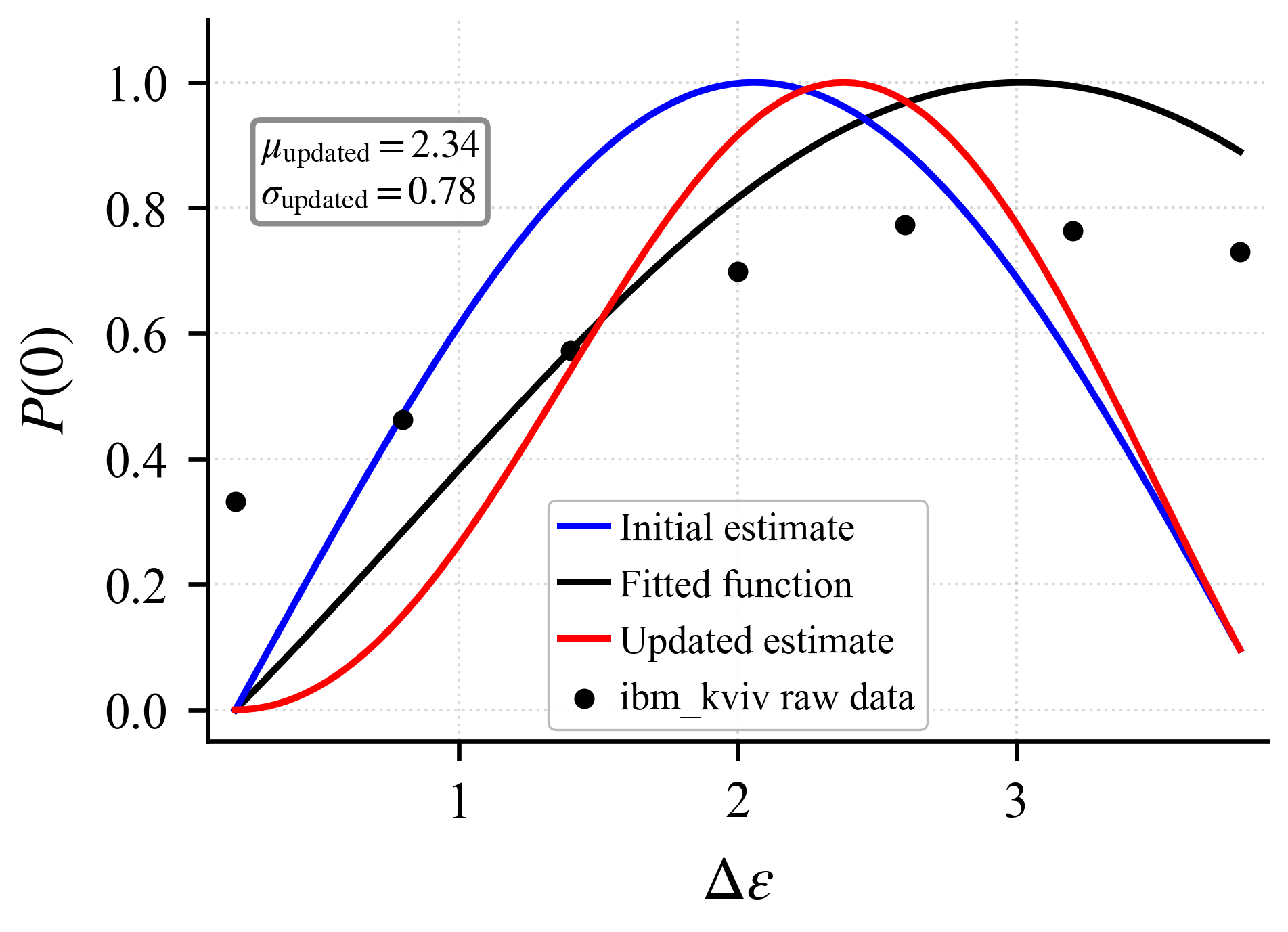}
}

\vspace{0.5cm}

\subfloat[$t = 1.6$\label{fig:triangular_frustration_e}]{%
    \includegraphics[width=0.45\textwidth]{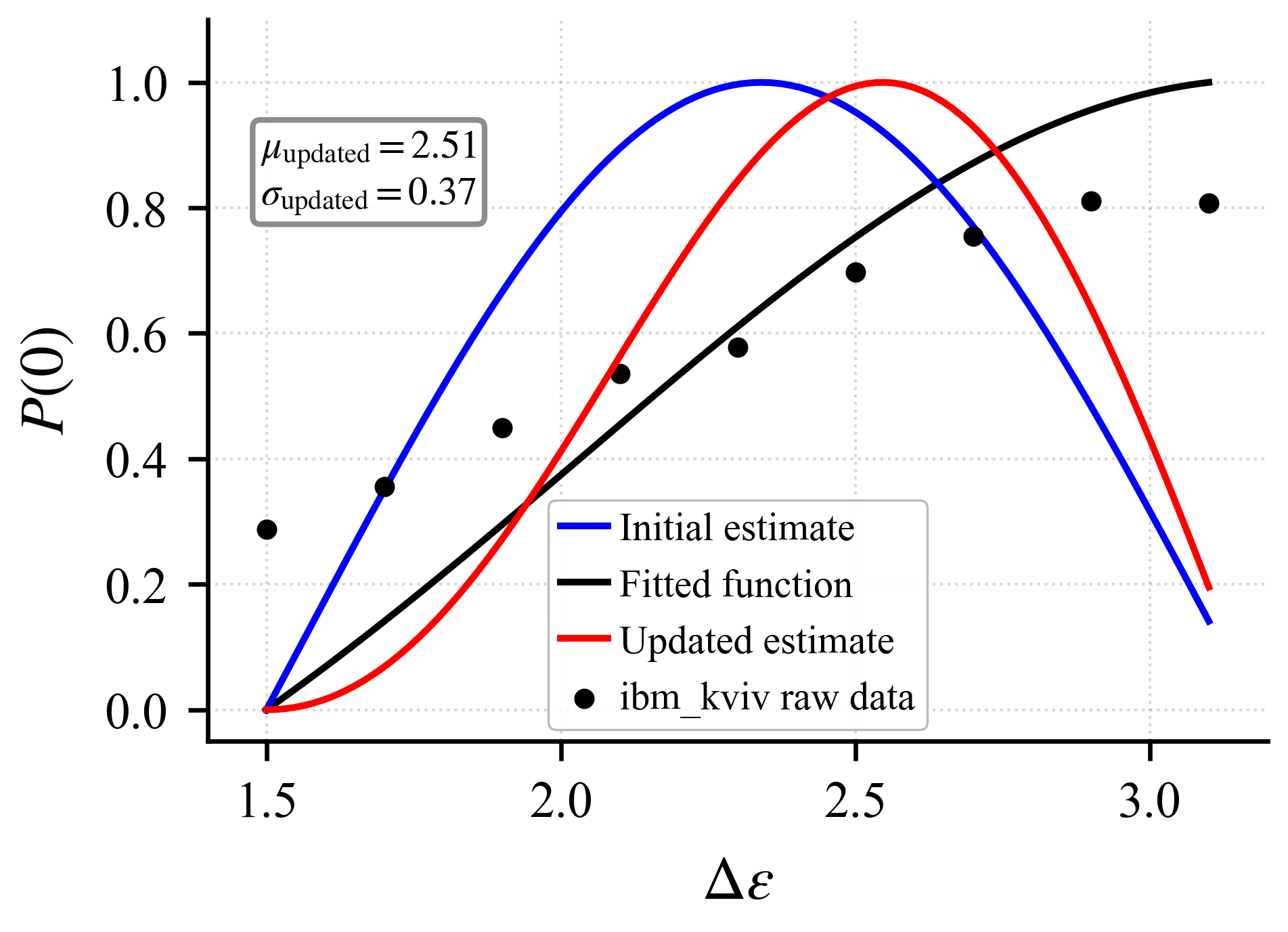}
}
\hfill
\subfloat[$t = 1.6$\label{fig:triangular_frustration_f}]{%
    \includegraphics[width=0.45\textwidth]{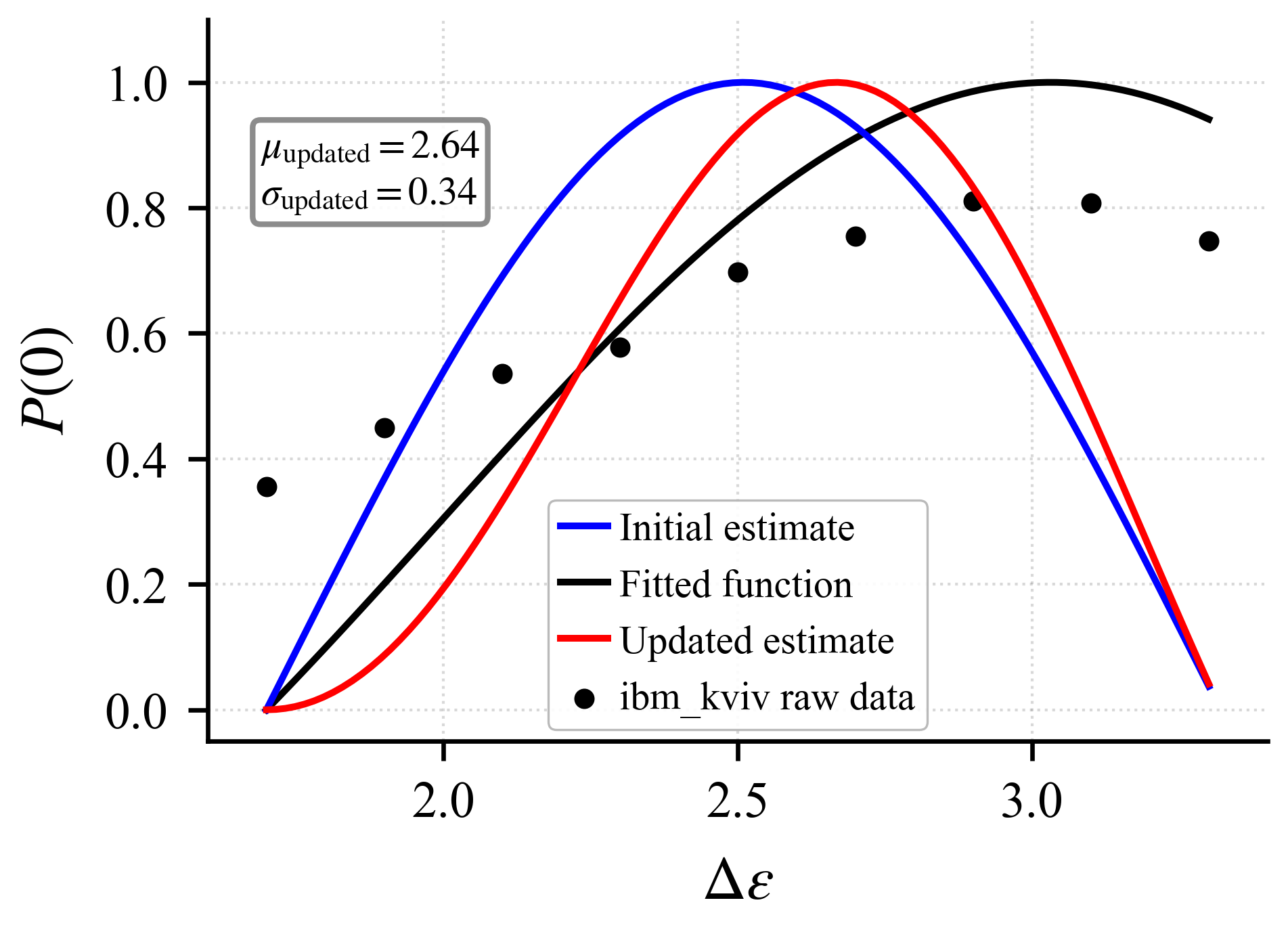}
}

\caption{ Initial estimates (blue), \texttt{ibm\_kyiv} raw data (black dots), fitted function (black), and updated 
 estimate (red) distributions showing progressive refinement of doublet-quartet energy gap in a frustrated three-spin triangle at different evolution times $t$. The updated means are: (a) $1.16 \pm 4.13$ at $t = 0.2$, (b) $1.76 \pm 1.75$ at $t = 0.4$, (c) $2.06 \pm 0.72$ at $t = 0.8$, (d) $2.34 \pm 0.78$ at $t = 0.8$, (e) $2.51 \pm 0.37$ at $t = 1.6$, and (f) $2.64 \pm 0.34$ at $t = 1.6$.}
\label{fig:updated_distributions_3spin_triangular_frustration}
\end{figure*}

\begin{figure*}[htbp]
    \centering
    \subfloat[$t=0.2$]{%
        \includegraphics[width=0.4\textwidth]{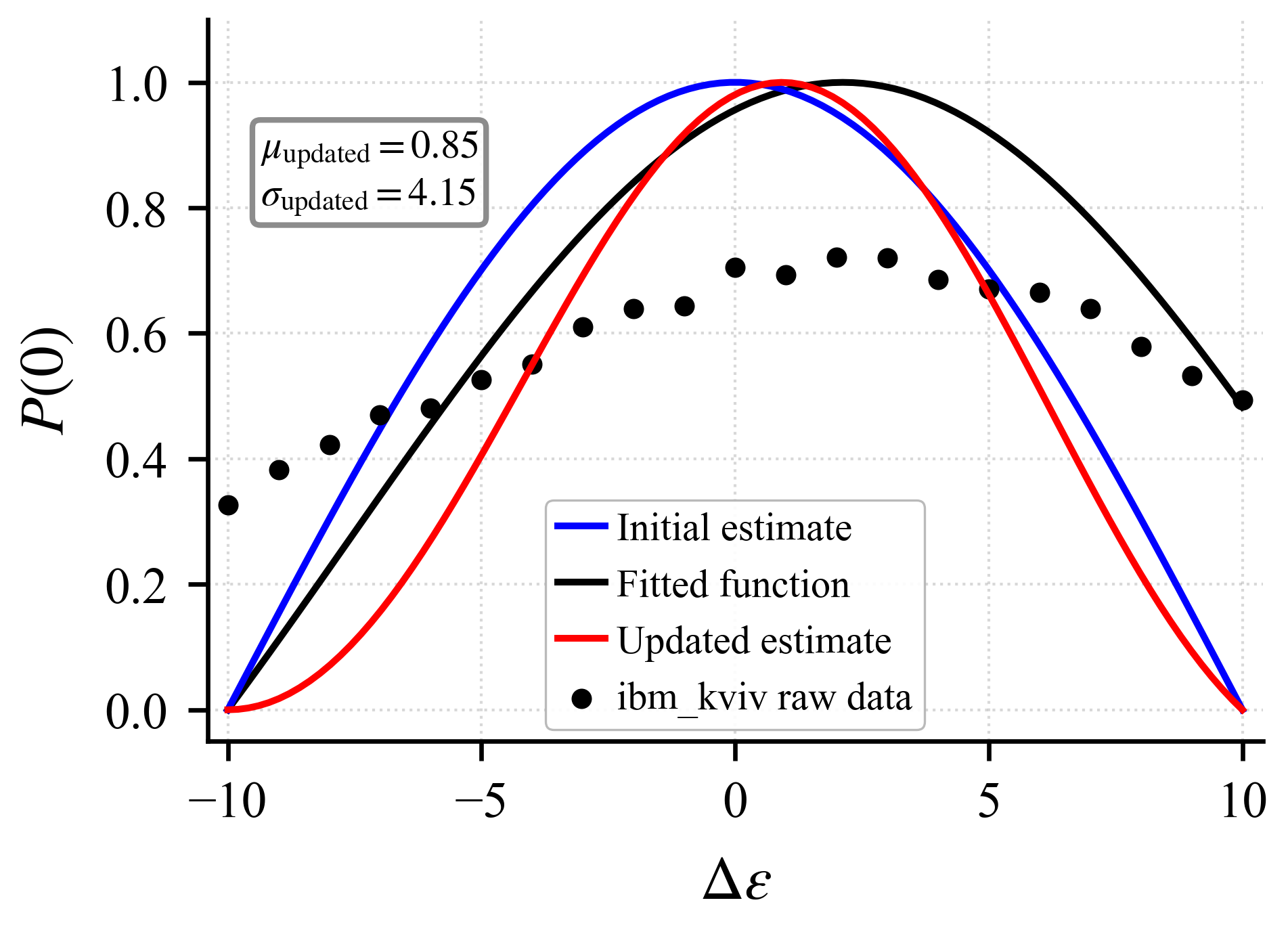}%
    }
    \subfloat[$t=0.4$]{%
        \includegraphics[width=0.4\textwidth]{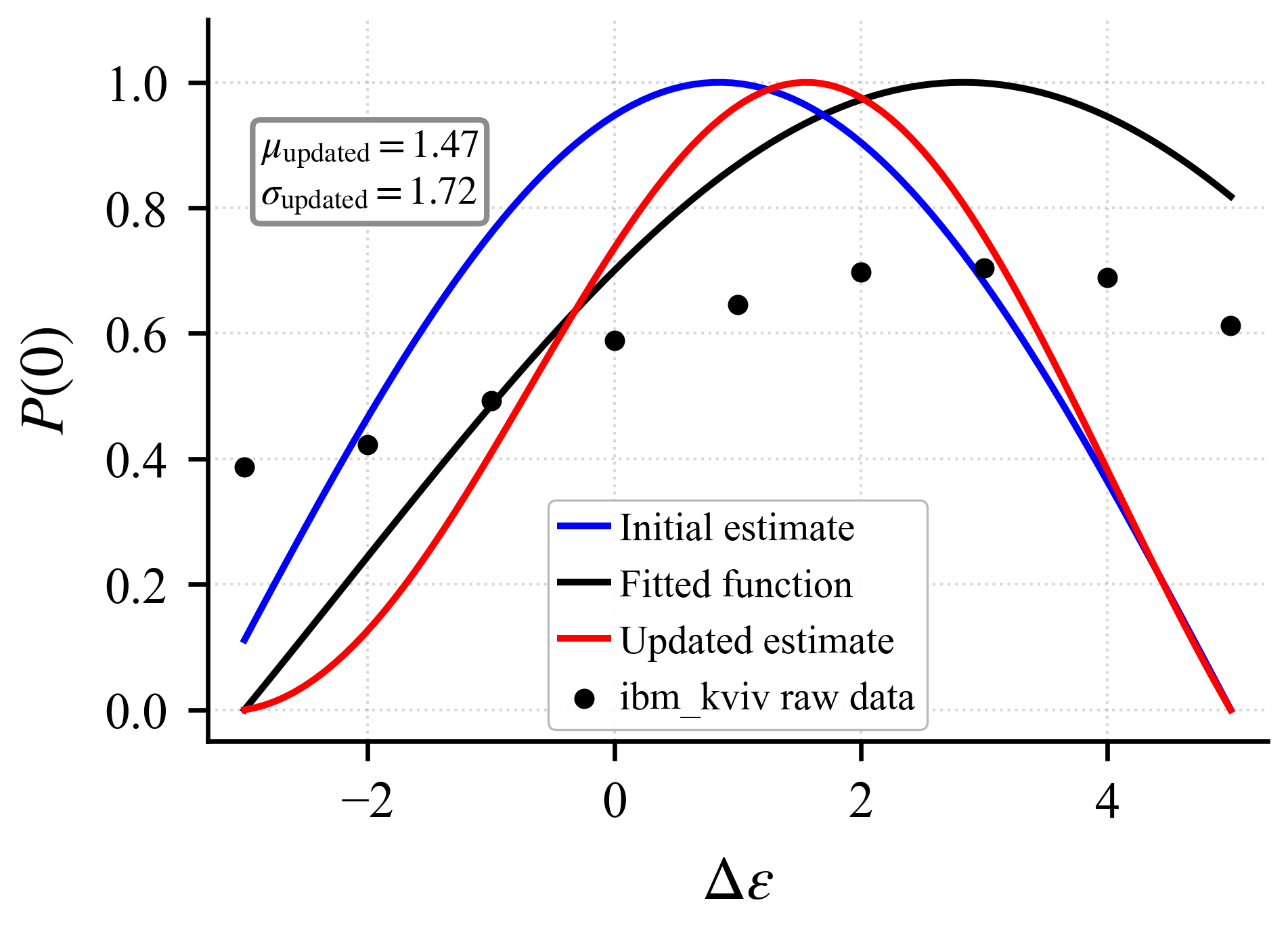}%
    }
    \subfloat[$t=0.6$]{%
        \includegraphics[width=0.4\textwidth]{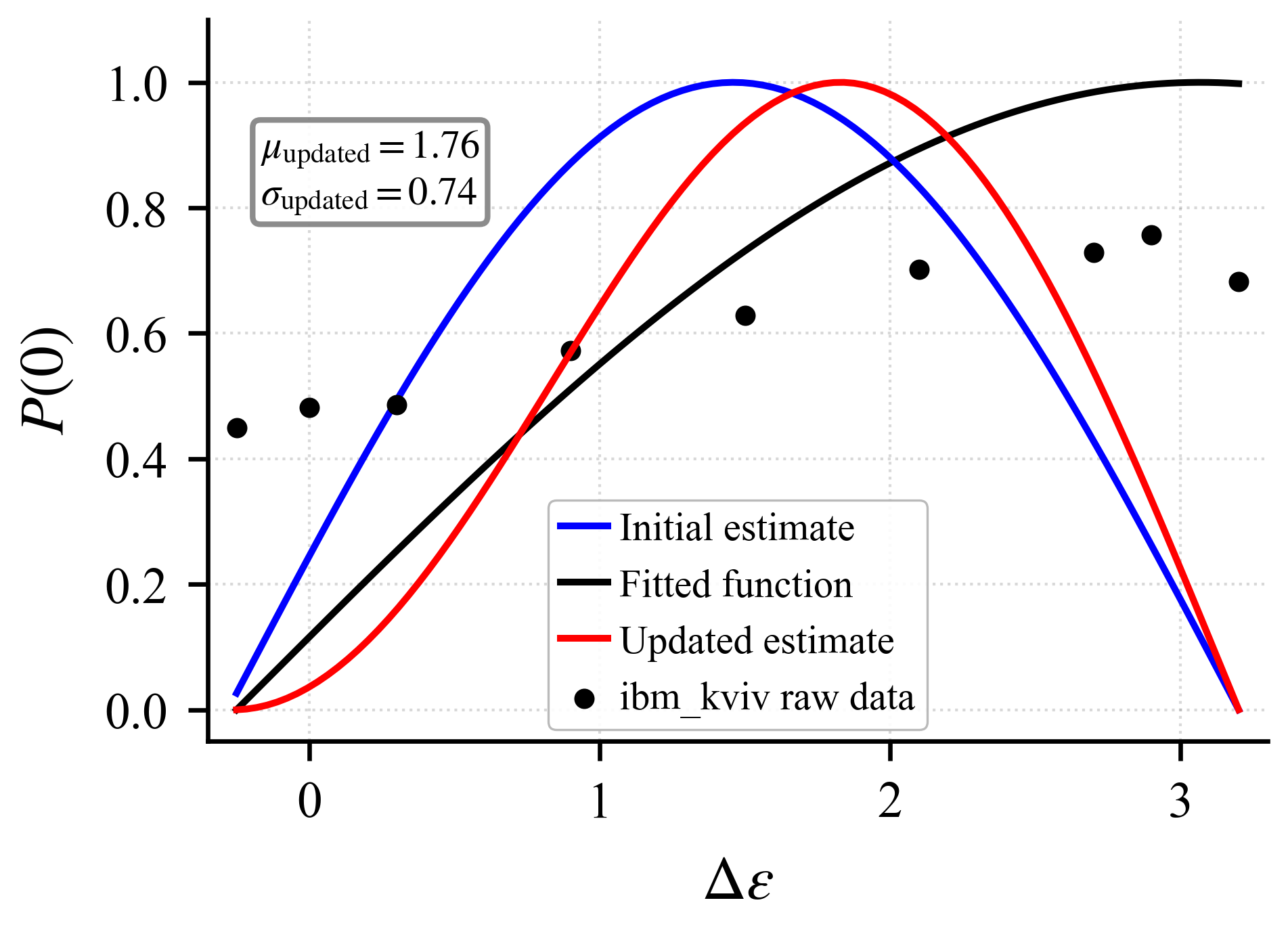}%
    }\\
    \subfloat[$t=0.6$]{%
        \includegraphics[width=0.4\textwidth]{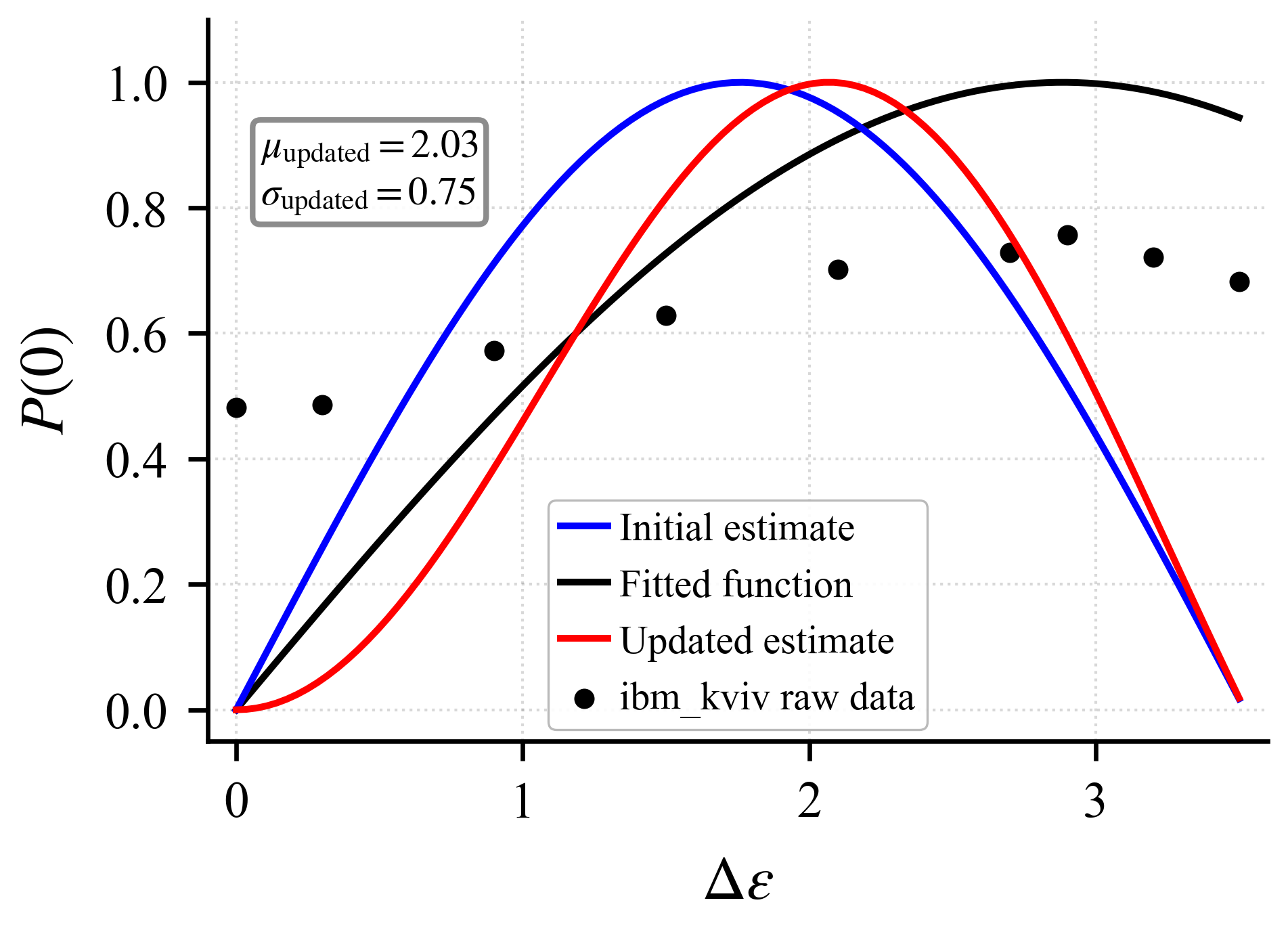}%
    }
    \subfloat[$t=0.6$]{%
        \includegraphics[width=0.4\textwidth]{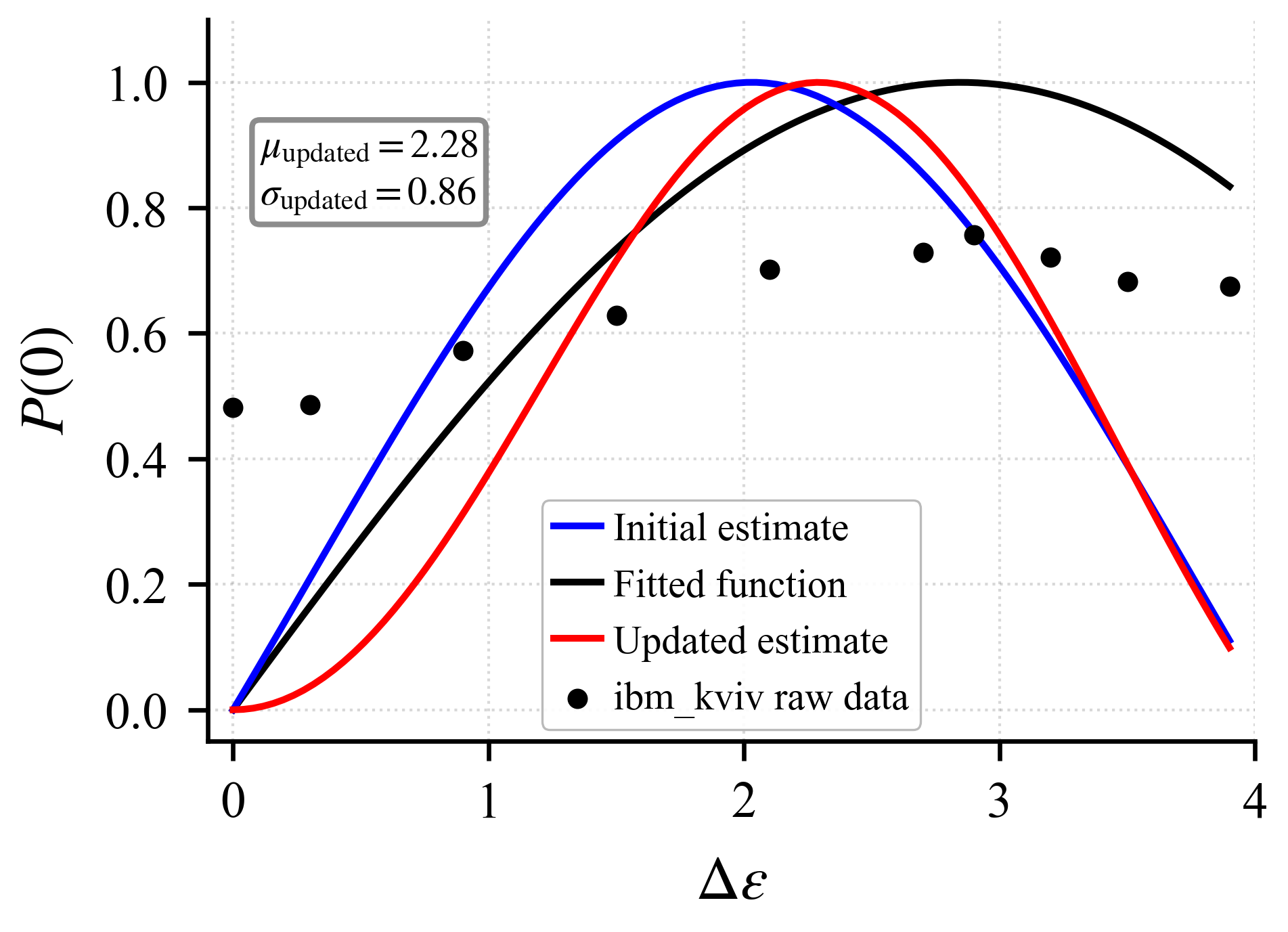}%
    }
    \subfloat[$t=1.6$]{%
        \includegraphics[width=0.4\textwidth]{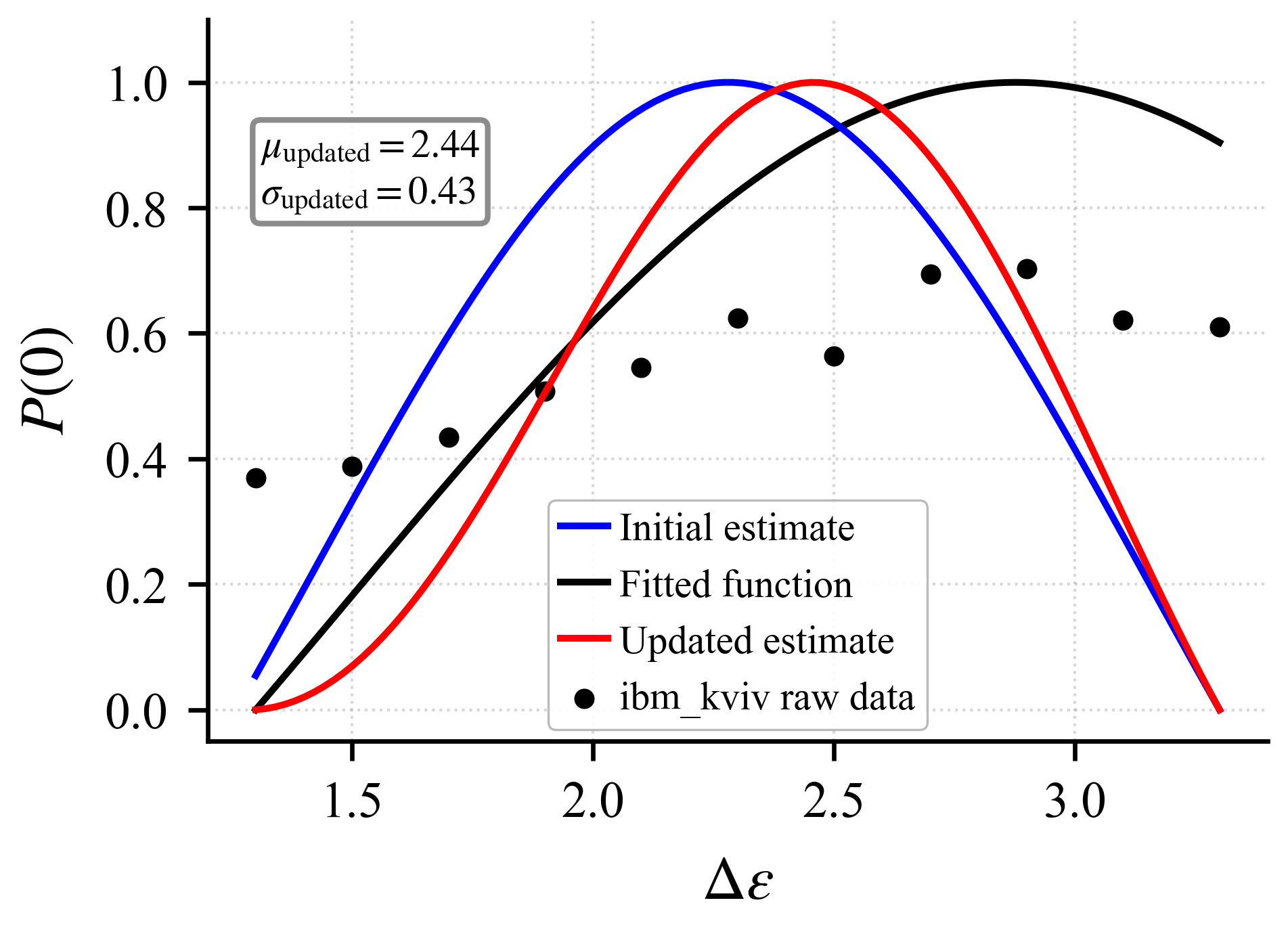}%
    }\\
    \subfloat[$t=1.6$]{%
        \includegraphics[width=0.4\textwidth]{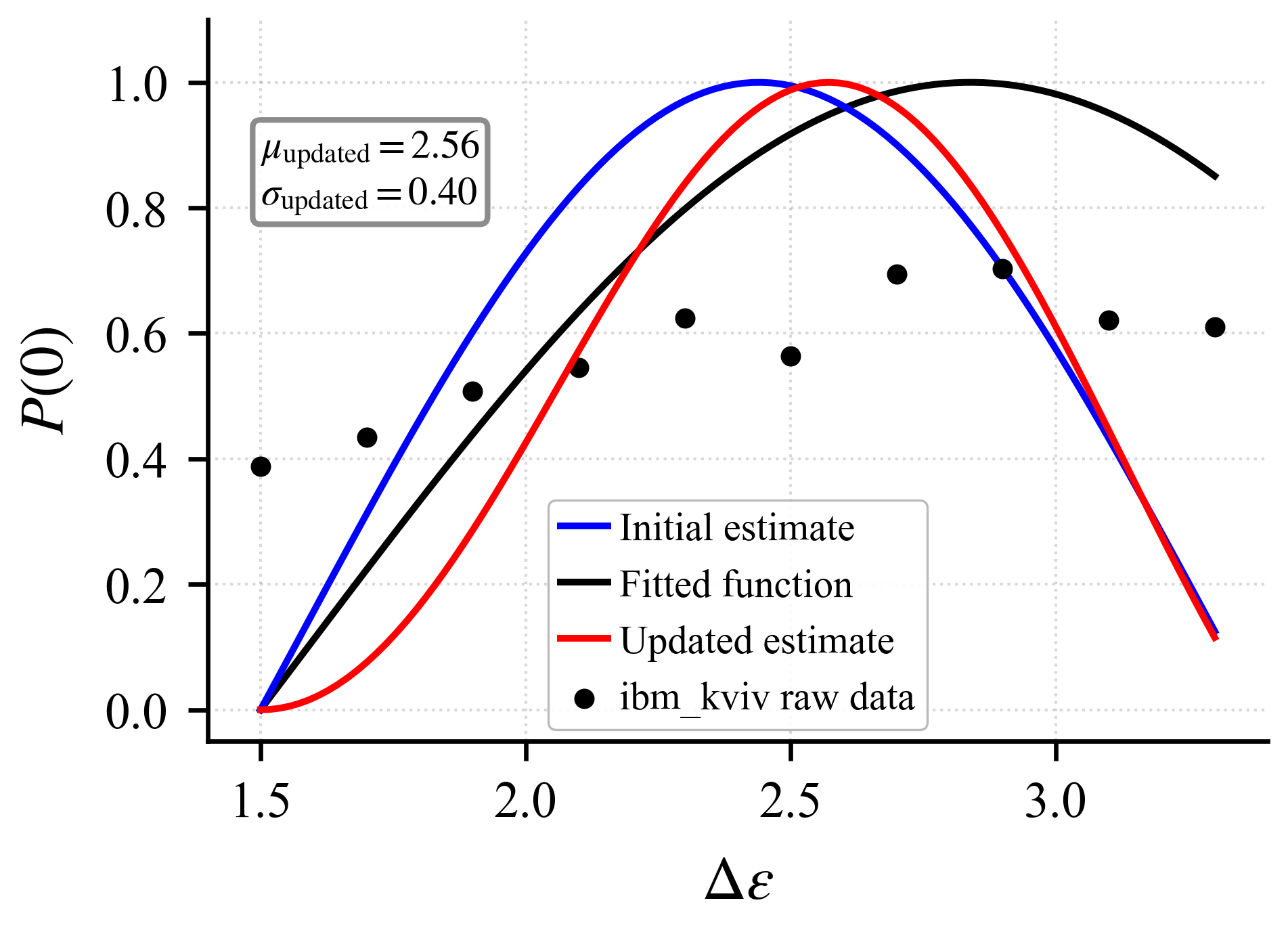}%
    }

    \caption{Time evolution of distributions, including initial estimates (blue), \texttt{ibm\_kyiv} raw data (black dots), fitted curves (black), and updated estimates (red), demonstrating the convergence of the QPDE algorithm for the first doublet state in a non-frustrated three-spin triangle. Updated means at different  evolution times $t$ are: (a) $0.85 \pm 4.15$ at $t = 0.2$, (b) $1.47 \pm 1.72$ at $t = 0.4$, (c) $1.76 \pm 0.74$ at $t = 0.6$, (d) $2.03 \pm 0.75$ at $t = 0.6$, (e) $2.28 \pm 0.86$ at $t = 0.6$, (f) $2.44 \pm 0.43$ at $t = 1.6$, and (g) $2.56 \pm 0.40$ at $t = 1.6$.}
    \label{fig:updated_distributions_without_frustration_doublet1}
\end{figure*}

\begin{figure*}[htbp!]
    \centering
    \subfloat[$t = 0.2$]{%
        \includegraphics[width=0.4\textwidth]{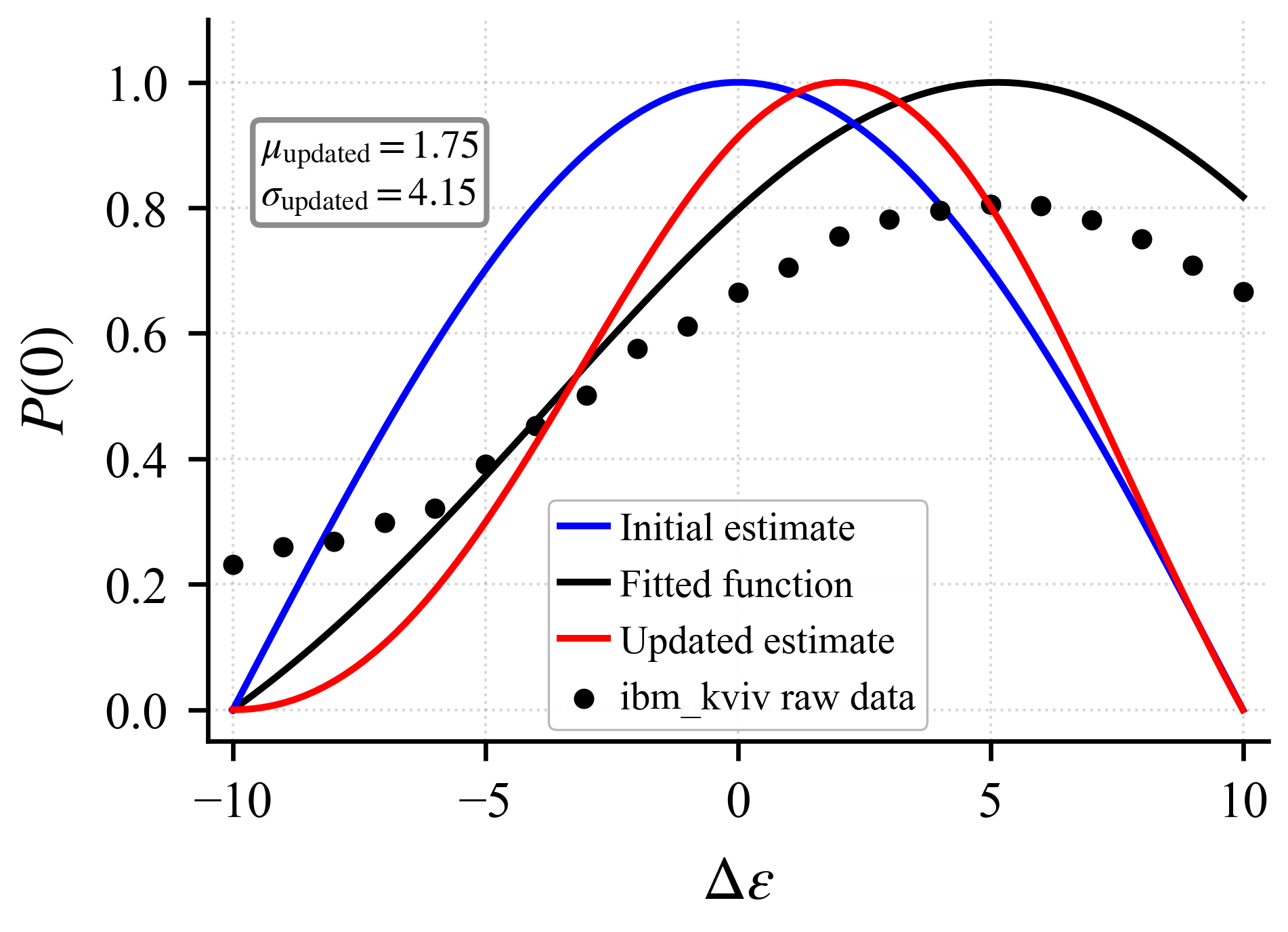}%
    }
    \subfloat[$t = 0.4$]{%
        \includegraphics[width=0.4\textwidth]{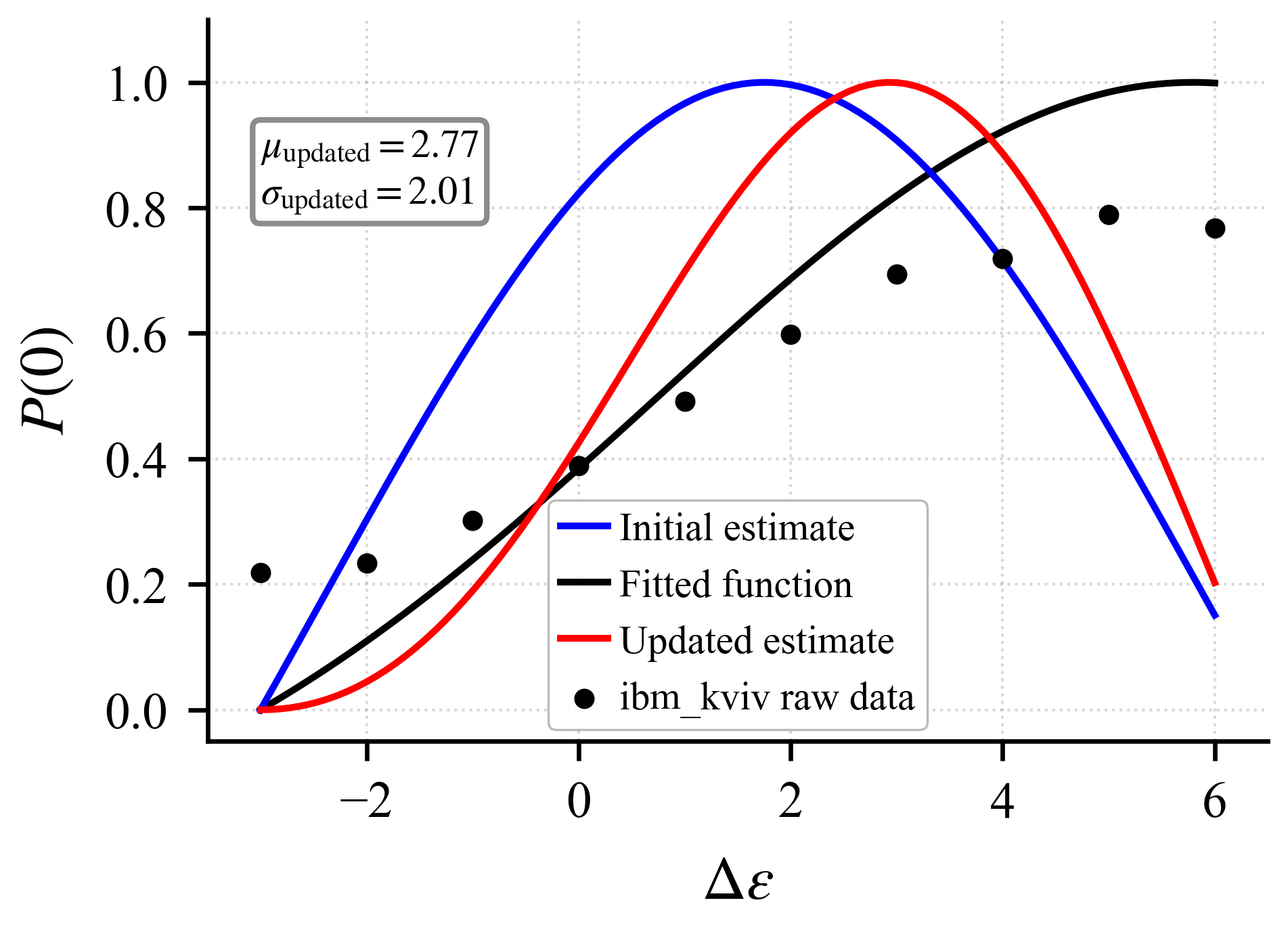}%
    }
    \subfloat[$t = 0.4$]{%
        \includegraphics[width=0.4\textwidth]{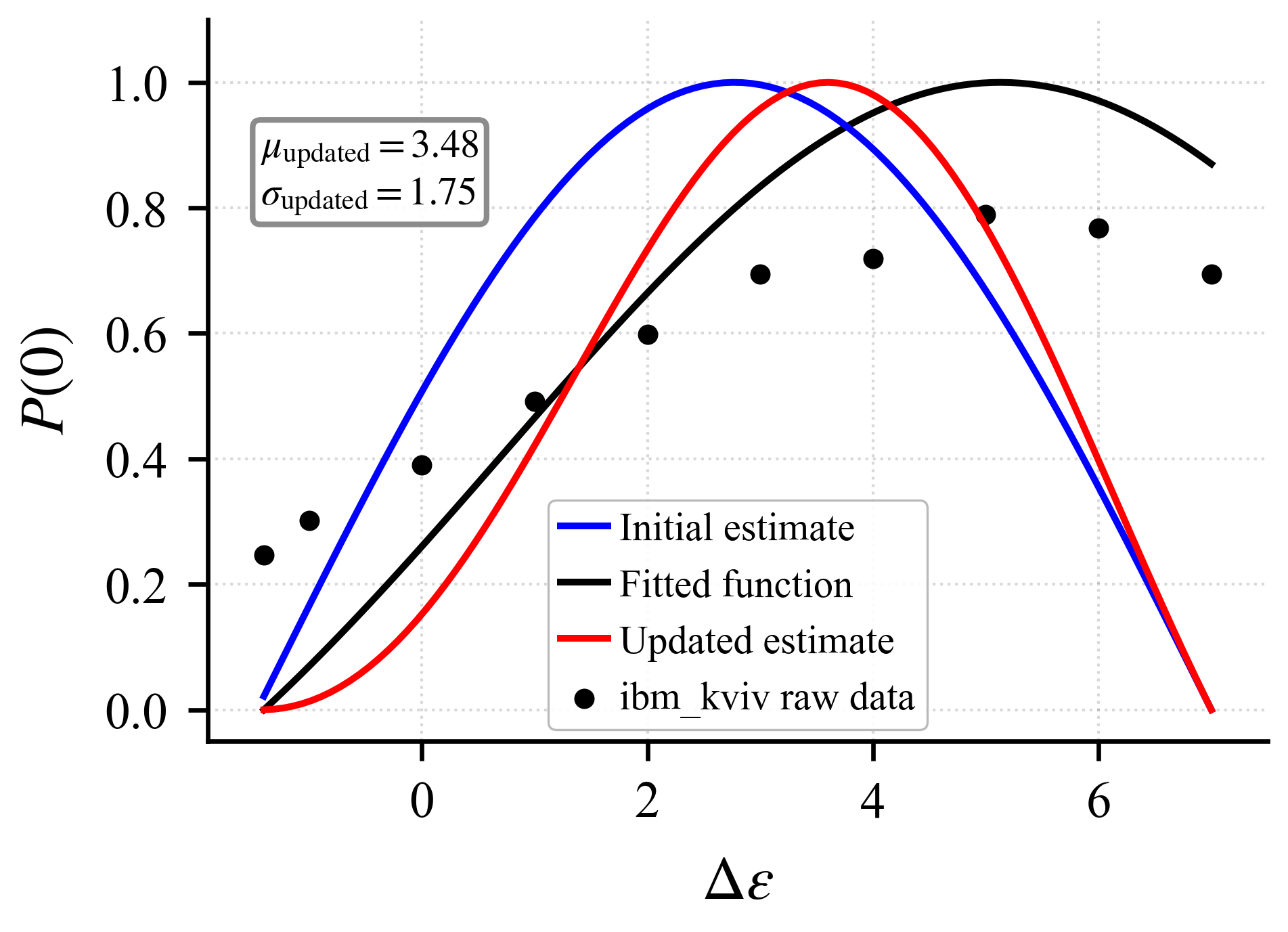}%
    }\\
    \subfloat[$t = 0.6$]{%
        \includegraphics[width=0.4\textwidth]{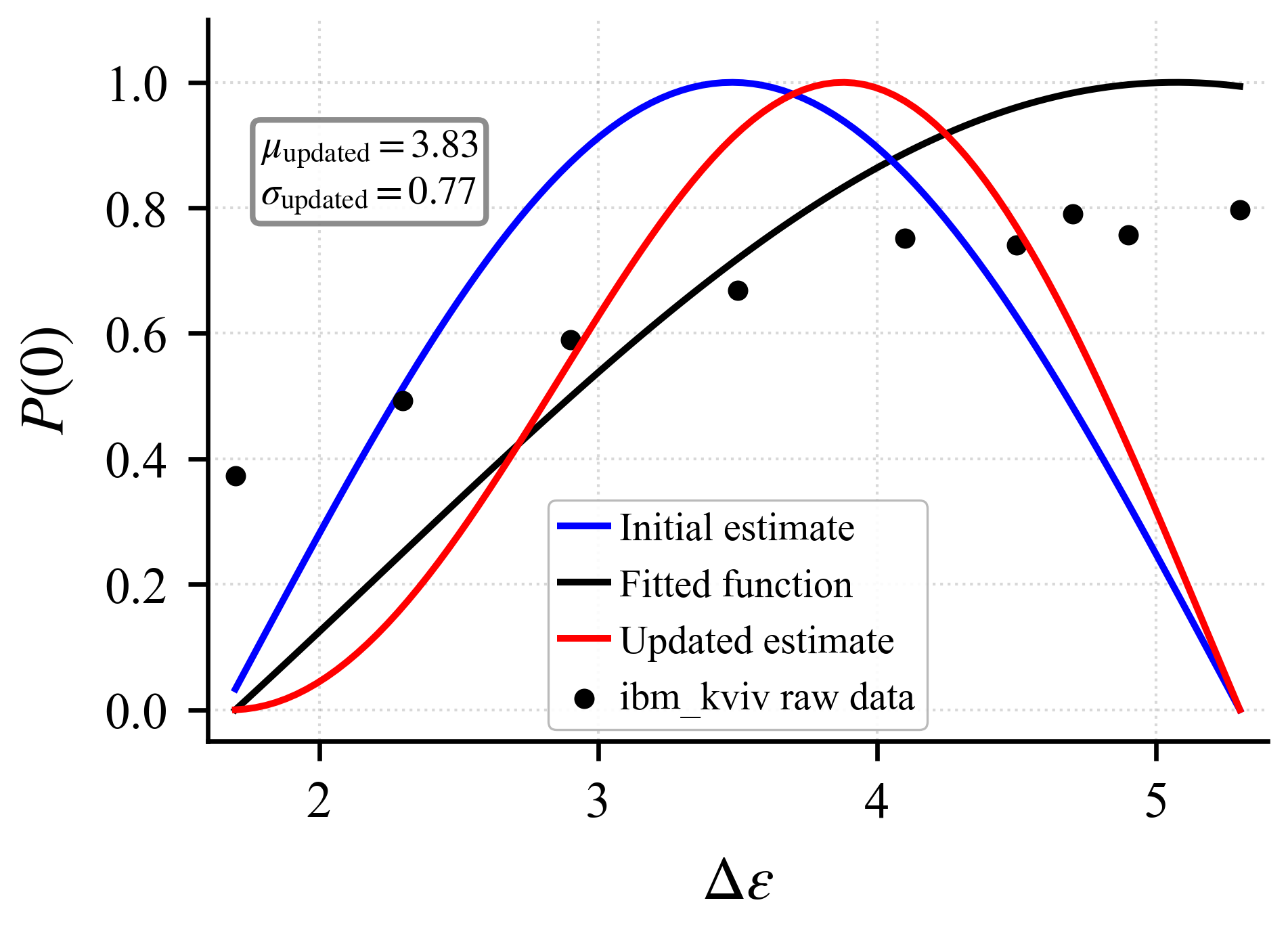}%
    }
    \subfloat[$t = 0.6$]{%
        \includegraphics[width=0.4\textwidth]{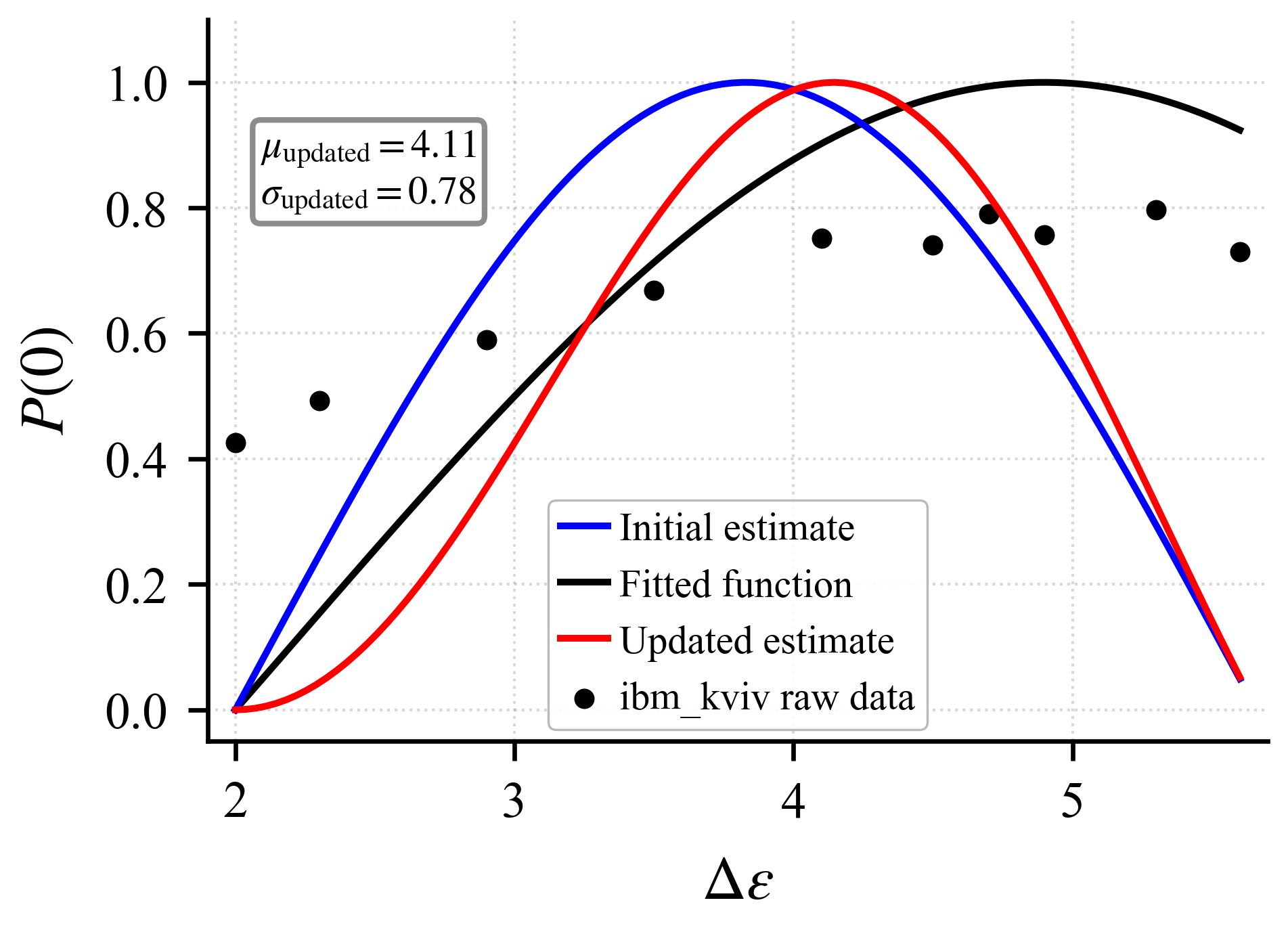}%
    }
    \subfloat[$t = 0.6$]{%
        \includegraphics[width=0.4\textwidth]{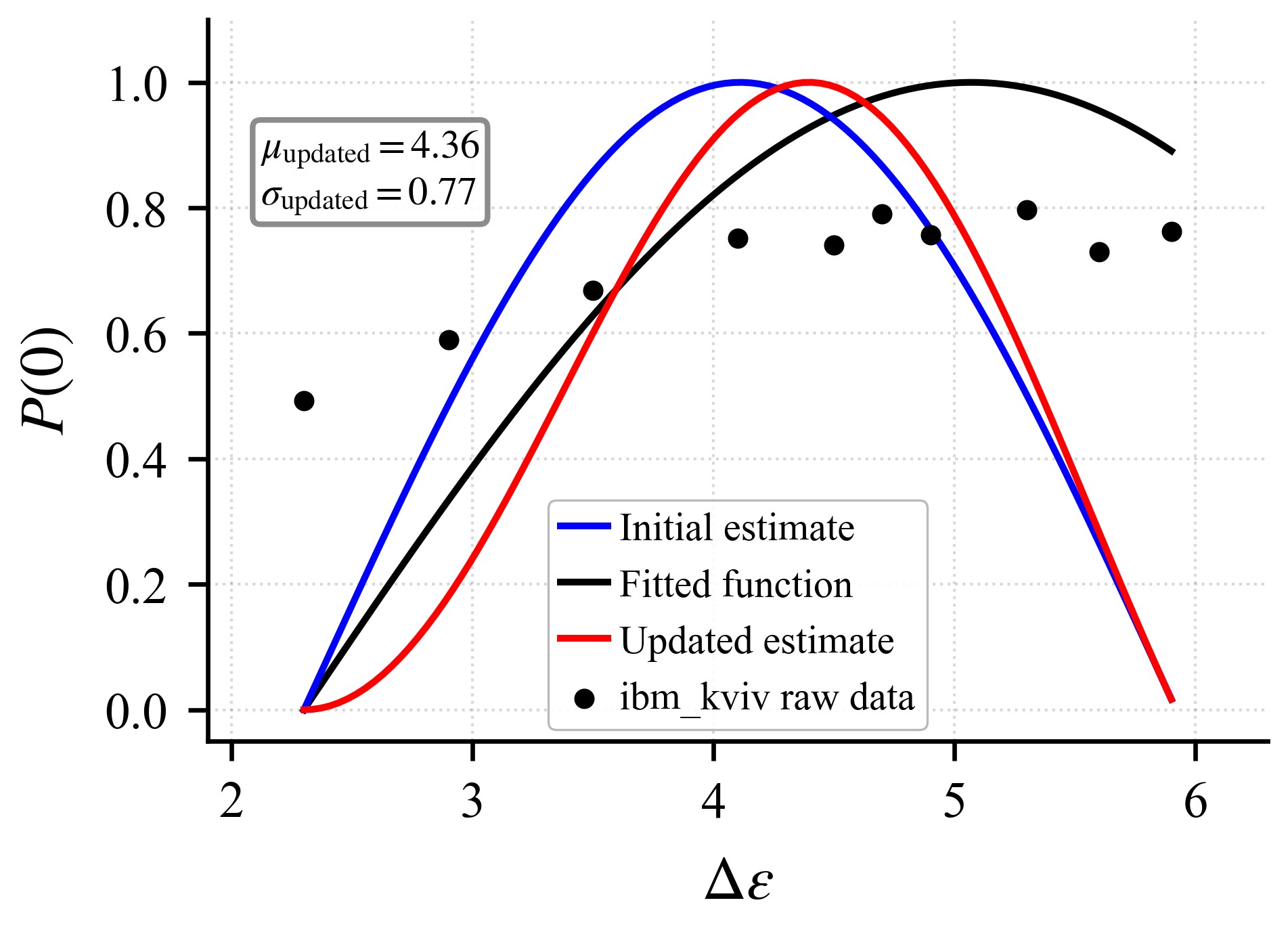}%
    }\\
    \subfloat[$t = 1.8$]{%
        \includegraphics[width=0.4\textwidth]{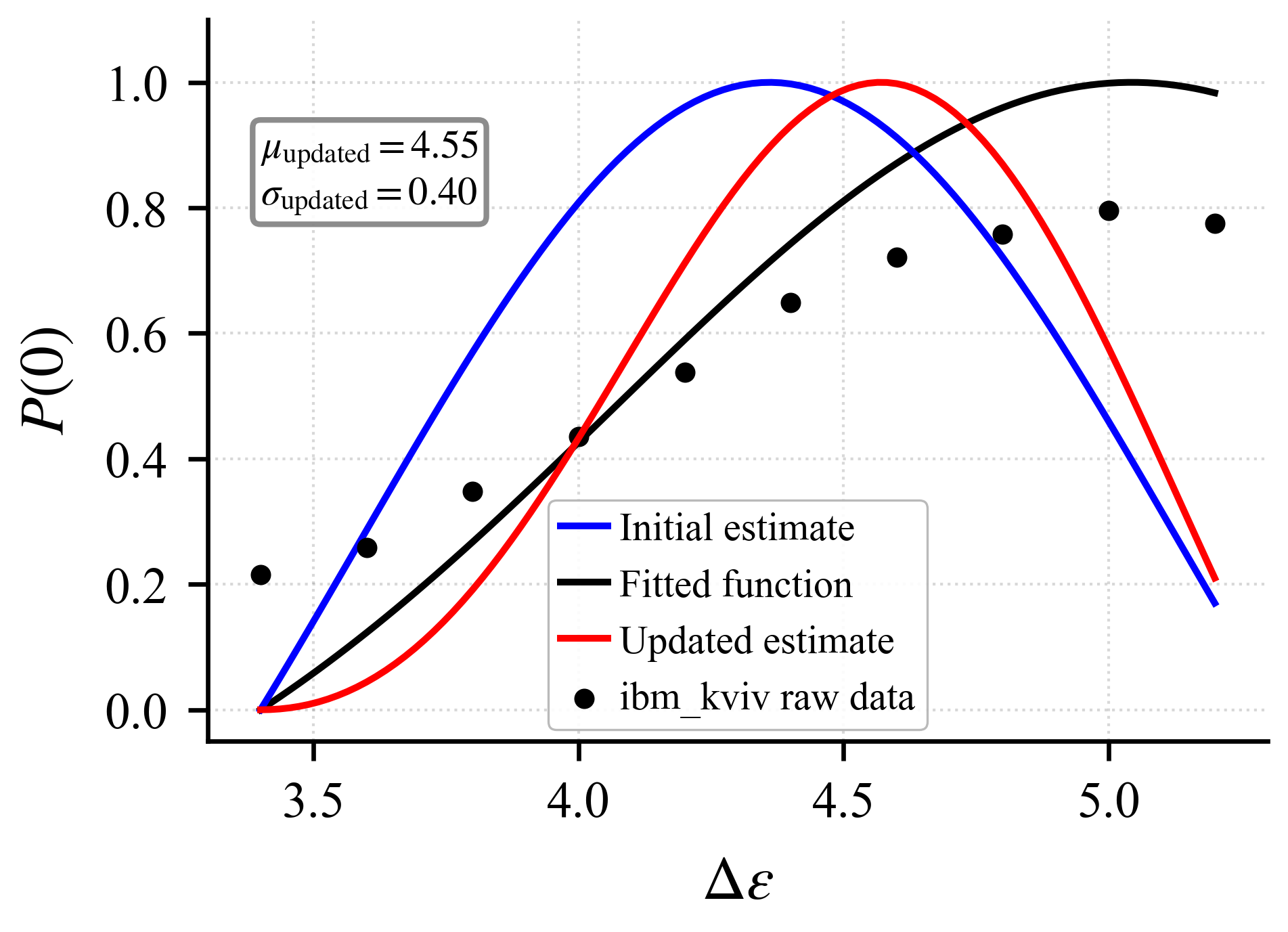}%
    }
    \subfloat[$t = 1.8$]{%
        \includegraphics[width=0.4\textwidth]{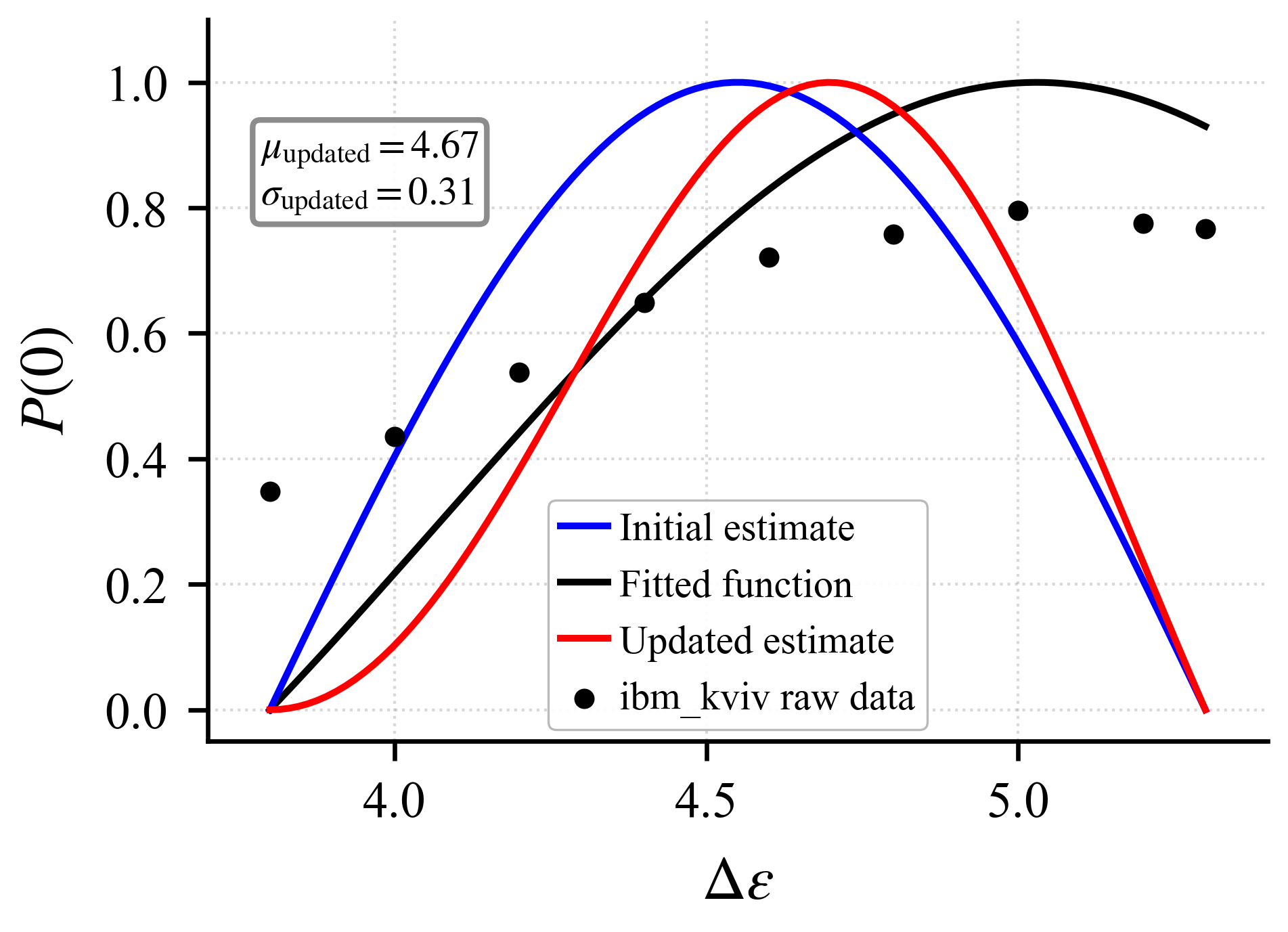}%
    }

    \caption{Initial estimates (blue), \texttt{ibm\_kyiv} raw data (black dots), fitted function (black), and updated estimates (red) illustrate the distributions in the QPDE algorithm for the second doublet state of a non-frustrated three-spin triangle. The updated means at different evolution times $t$ are: (a) $1.75 \pm 4.15$ at $t = 0.2$, (b) $2.77 \pm 2.01$ at $t = 0.4$, (c) $3.48 \pm 1.75$ at $t = 0.4$, (d) $3.83 \pm 0.77$ at $t = 0.6$, (e) $4.11 \pm 0.78$ at $t = 0.6$, (f) $4.36 \pm 0.77$ at $t = 0.6$, (g) $4.55 \pm 0.40$ at $t = 1.8$, and (h) $4.67 \pm 0.31$ at $t = 1.8$.}
    \label{fig:updated_distributions_without_frustration_doublet2}
\end{figure*}

\begin{figure*}[htbp!]
    \centering
    \begin{minipage}{0.4\textwidth}
        \centering
        \includegraphics[width=\textwidth]{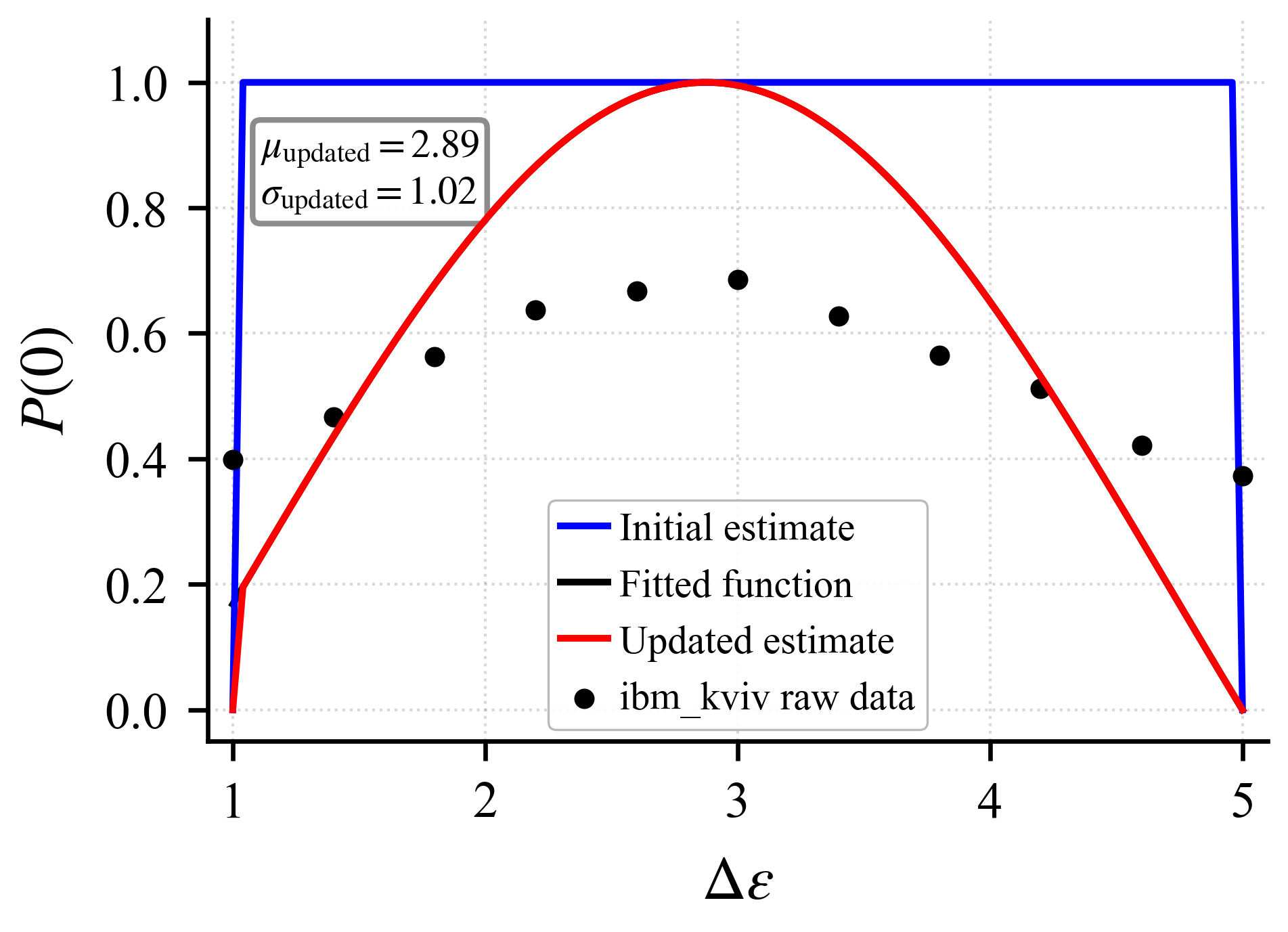}
        \caption{(a) $t = 1.2$}
    \end{minipage}
    \begin{minipage}{0.4\textwidth}
        \centering
        \includegraphics[width=\textwidth]{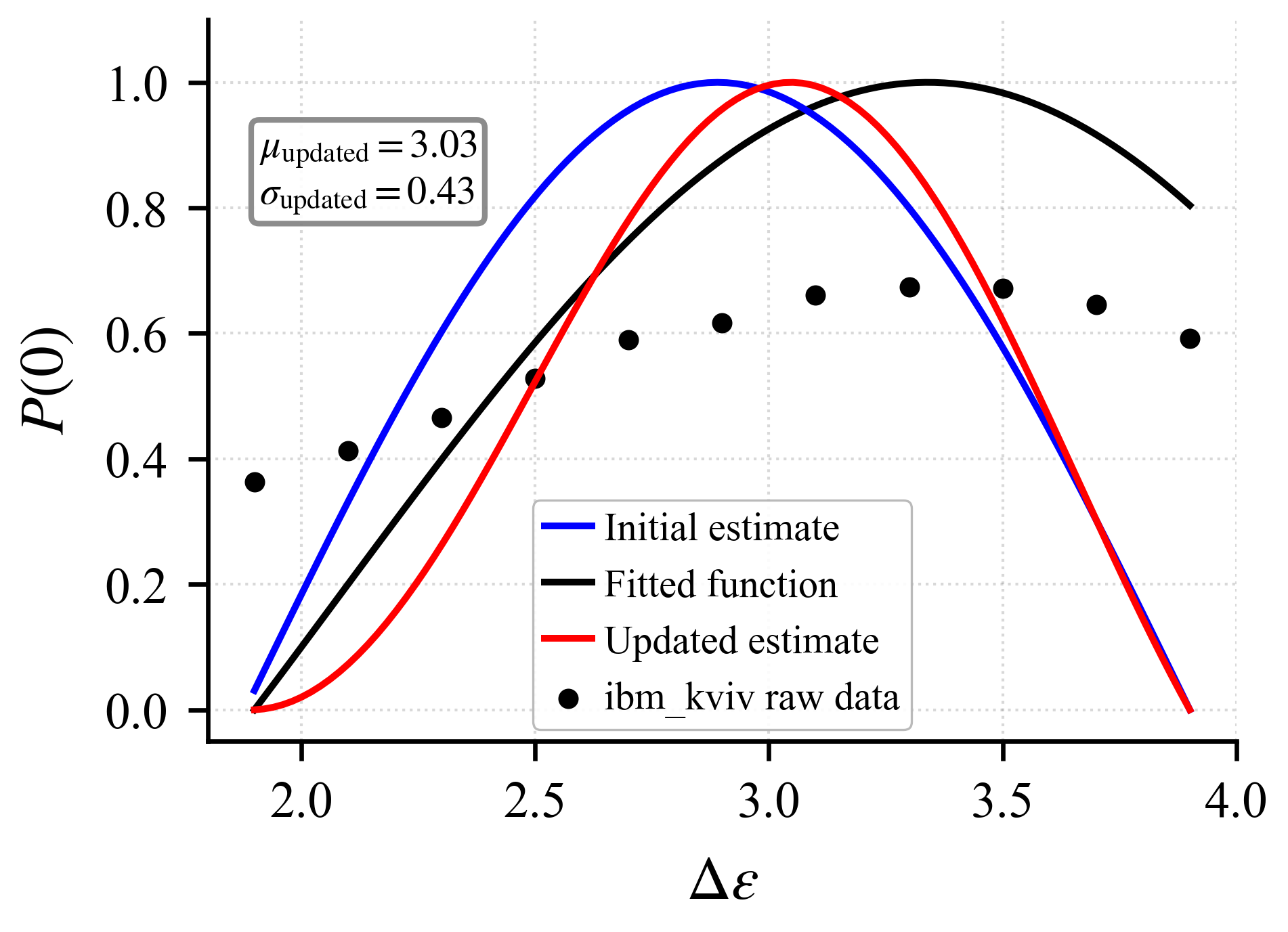}
        \caption{(b) $t = 1.8$}
    \end{minipage}
    \begin{minipage}{0.4\textwidth}
        \centering
        \includegraphics[width=\textwidth]{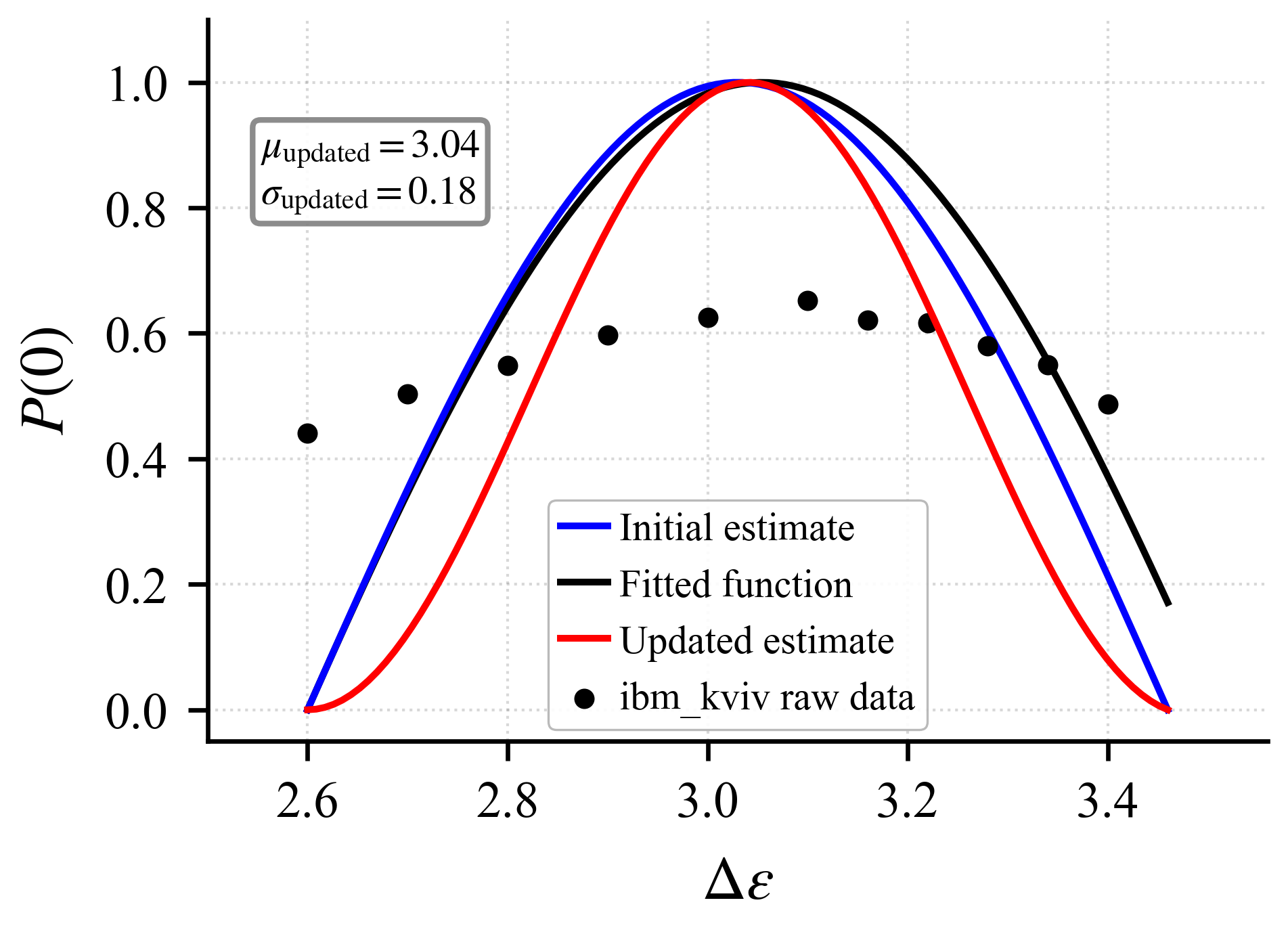}
        \caption{(c) $t = 4.2$}
    \end{minipage}
    
    \caption{Progression of initial (blue), \texttt{ibm\_kyiv} raw data (black dots), fitted (black), and updated (red) distributions for the QPDE algorithm applied to a three-spin asymmetric linear chain at various time points: (a) $t = 1.2$ with updated mean $2.89 \pm 1.02$, (b) $t = 1.8$ showing updated mean $3.03 \pm 0.43$, (c) $t = 4.2$ showing updated mean $3.04 \pm 0.18$.}
    \label{fig:updated_distributions_asymmetric}
\end{figure*}

\newpage
\section{Constant Depth Data Table}
\begin{table}[h]
\centering
\caption{Two spin system ($\Delta E_{S-T}$) data}
\begin{adjustbox}{max width=\textwidth}
\begin{tabular}{|c|c|c|c|c|c|c|c|}
\hline
\textbf{Time} & \textbf{\begin{tabular}[c]{@{}c@{}}Number of\\ Totter Steps\end{tabular}} & \multicolumn{2}{c|}{\textbf{Before Optimization}} & \multicolumn{2}{c|}{\textbf{After Optimization}} & \multicolumn{2}{c|}{\textbf{After ISA Circuit}} \\ \hline
 &  & \textbf{Circuit Depth} & \textbf{\begin{tabular}[c]{@{}c@{}}Number of 2\\ Qubit Gates\end{tabular}} & \textbf{Circuit Depth} & \textbf{\begin{tabular}[c]{@{}c@{}}Number of 2\\ Qubit Gates\end{tabular}} & \textbf{Circuit Depth} & \textbf{\begin{tabular}[c]{@{}c@{}}Number of 2\\ Qubit Gates\end{tabular}} \\ \hline
0.2 & 1 & 13 & 6 & 13 & 6 & 26 & 6 \\ 
0.4 & 1 & 13 & 6 & 13 & 6 & 26 & 6 \\ 
0.8 & 1 & 13 & 6 & 13 & 6 & 26 & 6 \\ 
2.4 & 1 & 13 & 6 & 13 & 6 & 26 & 6 \\ 
\hline
\end{tabular}
\end{adjustbox}
\end{table}

\begin{table}[h]
\centering
\caption{Three-spin linear chain ($\Delta E_{D_2-Q}$) data}
\begin{adjustbox}{max width=\textwidth}
\begin{tabular}{|c|c|c|c|c|c|c|c|}
\hline
\textbf{Time} & \textbf{\begin{tabular}[c]{@{}c@{}}Number of\\ Totter Steps\end{tabular}} & \multicolumn{2}{c|}{\textbf{Before Optimization}} & \multicolumn{2}{c|}{\textbf{After Optimization}} & \multicolumn{2}{c|}{\textbf{After ISA Circuit}} \\ \hline
 &  & \textbf{Circuit Depth} & \textbf{\begin{tabular}[c]{@{}c@{}}Number of 2\\ Qubit Gates\end{tabular}} & \textbf{Circuit Depth} & \textbf{\begin{tabular}[c]{@{}c@{}}Number of 2\\ Qubit Gates\end{tabular}} & \textbf{Circuit Depth} & \textbf{\begin{tabular}[c]{@{}c@{}}Number of 2\\ Qubit Gates\end{tabular}} \\ \hline
0.2 & 30 & 368 & 186 & 54 & 22 & 184 & 46 \\
0.6 & 90 & 1088 & 546 & 54 & 22 & 184 & 46 \\
1.0 & 150 & 1808 & 906 & 54 & 22 & 184 & 46 \\
4.2 & 620 & 7448 & 3726 & 54 & 22 & 184 & 46 \\ \hline
\end{tabular}
\end{adjustbox}
\end{table}

\begin{table}[h]
\centering
\caption{Three-spin triangle without frustration ($\Delta E_{D_1-Q}$) data}
\begin{adjustbox}{max width=\textwidth}
\begin{tabular}{|c|c|c|c|c|c|c|c|}
\hline
\textbf{Time} & \textbf{\begin{tabular}[c]{@{}c@{}}Number of\\ Totter Steps\end{tabular}} & \multicolumn{2}{c|}{\textbf{Before Optimization}} & \multicolumn{2}{c|}{\textbf{After Optimization}} & \multicolumn{2}{c|}{\textbf{After ISA Circuit}} \\ \hline
 &  & \textbf{Circuit Depth} & \textbf{\begin{tabular}[c]{@{}c@{}}Number of 2\\ Qubit Gates\end{tabular}} & \textbf{Circuit Depth} & \textbf{\begin{tabular}[c]{@{}c@{}}Number of 2\\ Qubit Gates\end{tabular}} & \textbf{Circuit Depth} & \textbf{\begin{tabular}[c]{@{}c@{}}Number of 2\\ Qubit Gates\end{tabular}} \\ \hline
0.2 & 30 & 529 & 286 & 78 & 34 & 247 & 54 \\
0.4 & 60 & 1039 & 556 & 78 & 34 & 247 & 54 \\
0.6 & 90 & 1549 & 826 & 78 & 34 & 247 & 54 \\
1.6 & 210 & 3589 & 1906 & 78 & 34 & 247 & 54 \\
\hline
\end{tabular}
\end{adjustbox}
\end{table}

\begin{table}[h]
\centering
\caption{Three-spin triangle without frustration ($\Delta E_{D_2-Q}$) data}
\begin{adjustbox}{max width=\textwidth}
\begin{tabular}{|c|c|c|c|c|c|c|c|}
\hline
\textbf{Time} & \textbf{\begin{tabular}[c]{@{}c@{}}Number of\\ Totter Steps\end{tabular}} & \multicolumn{2}{c|}{\textbf{Before Optimization}} & \multicolumn{2}{c|}{\textbf{After Optimization}} & \multicolumn{2}{c|}{\textbf{After ISA Circuit}} \\ \hline
 &  & \textbf{Circuit Depth} & \textbf{\begin{tabular}[c]{@{}c@{}}Number of 2\\ Qubit Gates\end{tabular}} & \textbf{Circuit Depth} & \textbf{\begin{tabular}[c]{@{}c@{}}Number of 2\\ Qubit Gates\end{tabular}} & \textbf{Circuit Depth} & \textbf{\begin{tabular}[c]{@{}c@{}}Number of 2\\ Qubit Gates\end{tabular}} \\ \hline
0.2 & 30 & 521 & 278 & 64 & 22 & 245 & 47 \\
0.4 & 60 & 1031 & 548 & 64 & 22 & 245 & 47 \\
0.6 & 90 & 1541 & 818 & 64 & 22 & 245 & 47 \\
1.8 & 270 & 4601 & 2438 & 64 & 22 & 245 & 47 \\
 \hline
\end{tabular}
\end{adjustbox}
\end{table}

\begin{table}[h]
\centering
\caption{Three-spin triangle with frustration ($\Delta E_{D_2-Q}$) data}
\begin{adjustbox}{max width=\textwidth}
\begin{tabular}{|c|c|c|c|c|c|c|c|}
\hline
\textbf{Time} & \textbf{\begin{tabular}[c]{@{}c@{}}Number of\\ Totter Steps\end{tabular}} & \multicolumn{2}{c|}{\textbf{Before Optimization}} & \multicolumn{2}{c|}{\textbf{After Optimization}} & \multicolumn{2}{c|}{\textbf{After ISA Circuit}} \\ \hline
 &  & \textbf{Circuit Depth} & \textbf{\begin{tabular}[c]{@{}c@{}}Number of 2\\ Qubit Gates\end{tabular}} & \textbf{Circuit Depth} & \textbf{\begin{tabular}[c]{@{}c@{}}Number of 2\\ Qubit Gates\end{tabular}} & \textbf{Circuit Depth} & \textbf{\begin{tabular}[c]{@{}c@{}}Number of 2\\ Qubit Gates\end{tabular}} \\ \hline
0.2 & 30 & 521 & 276 & 55 & 22 & 227 & 47 \\
0.4 & 60 & 1031 & 456 & 55 & 22 & 227 & 47 \\
0.8 & 120 & 2051 & 1086 & 55 & 22 & 227 & 47 \\
1.6 & 240 & 4091 & 2166 & 55 & 22 & 227 & 47 \\ \hline
\end{tabular}
\end{adjustbox}
\end{table}

\begin{table}[h]
\centering
\caption{Three-spin asymmetric chain ($\Delta E_{D_1-Q}$) data}
\begin{adjustbox}{max width=\textwidth}
\begin{tabular}{|c|c|c|c|c|c|c|c|}
\hline
\textbf{Time} & \textbf{\begin{tabular}[c]{@{}c@{}}Number of\\ Totter Steps\end{tabular}} & \multicolumn{2}{c|}{\textbf{Before Optimization}} & \multicolumn{2}{c|}{\textbf{After Optimization}} & \multicolumn{2}{c|}{\textbf{After ISA Circuit}} \\ \hline
 &  & \textbf{Circuit Depth} & \textbf{\begin{tabular}[c]{@{}c@{}}Number of 2\\ Qubit Gates\end{tabular}} & \textbf{Circuit Depth} & \textbf{\begin{tabular}[c]{@{}c@{}}Number of 2\\ Qubit Gates\end{tabular}} & \textbf{Circuit Depth} & \textbf{\begin{tabular}[c]{@{}c@{}}Number of 2\\ Qubit Gates\end{tabular}} \\ \hline
1.2 & 180 & 3071 & 1626 & 57 & 21 & 177 & 37 \\
1.8 & 300 & 5111 & 2706 & 57 & 21 & 177 & 37 \\
4.2 & 620 & 10551 & 5586 & 57 & 21 & 177 & 37 \\ \hline
\end{tabular}
\end{adjustbox}
\end{table}